\documentclass[aps,pra,twocolumn,amsfonts,amssymb,amsmath,showpacs,
floatfix,nofootinbib,citesort]{revtex4-2}
\usepackage{mathrsfs}
\usepackage{amsfonts}
\usepackage{amstext}
\usepackage{amsmath}
\usepackage{amssymb}
\usepackage{bm}
\usepackage{CJK}
\usepackage{subfigure}
\usepackage{bbm}
\usepackage[dvips]{graphicx}
\def\qed{\leavevmode\unskip\penalty9999 \hbox{}\nobreak\hfill
     \quad\hbox{\leavevmode  \hbox to.77778em{%
              \hfil\vrule   \vbox to.675em%
               {\hrule width.6em\vfil\hrule}\vrule\hfil}}
     \par\vskip3pt}

\usepackage{amssymb}
\usepackage{graphicx}
\usepackage{graphics}
\usepackage{amsmath}
\usepackage{amsthm}
\usepackage{color}
\usepackage{dsfont}
\usepackage{textcomp}
\usepackage{footnote}
\definecolor{darkred}  {rgb}{0.5,0,0}
\definecolor{darkblue} {rgb}{0,0,0.5}
\definecolor{darkgreen}{rgb}{0,0.5,0}
\usepackage{hyperref}
\hypersetup{
	pdftitle = {QRT Proposal},
	pdfauthor = {},
	colorlinks = true,
	urlcolor  = blue,         
	linkcolor = red,     
	citecolor = blue,    
	filecolor = darkred       
}
\usepackage{mathtools}
\def\ra{\rangle}
\def\la{\langle}

\def\ot{\otimes}
\newtheorem{theorem}{Theorem}

\newtheorem{pro}{Proposition}

\newcommand{\bea}{\begin{eqnarray}}
\newcommand{\eea}{\end{eqnarray}}
\newcommand{\be}{\begin{equation}}
\newcommand{\ee}{\end{equation}}
\newcommand{\ba}{\begin{equation}\begin{aligned}}
\newcommand{\ea}{\end{aligned}\end{equation}}

\newcommand{\beax}{\begin{eqnarray*}}
\newcommand{\eeax}{\end{eqnarray*}}
\newcommand{\bex}{\begin{equation*}}
\newcommand{\eex}{\end{equation*}}

\theoremstyle{remark}

\def\be{\begin{equation}}
\def\ee{\end{equation}}

\newcommand{\mF}{\mathcal{F}}
\newcommand{\mE}{\mathcal{E}}
\newcommand{\mC}{\mathcal{C}}

\newcommand{\mH}{\mathcal{H}}

\newcommand{\mN}{\mathcal{N}}

\newcommand{\mS}{\mathcal{S}}

\newcommand{\lr}{\rangle\langle}

\newcommand{\tr}{{\rm Tr}}

\newcommand{\mbb}[1]{\mathbb{#1}}

\begin{document}


\preprint{APS/123-QED}
\begin{CJK*}{GB}{gbsn}
\title{Complete genuine multipartite entanglement monotone \\}


\author{Yu Guo}
\email{guoyu3@aliyun.com}

\affiliation{School of Mathematical Sciences, Inner Mongolia University, Hohhot, Inner Mongolia 010021, People's Republic of China}



\begin{abstract}
A complete characterization and quantification of entanglement, particularly the multipartite entanglement, remains an unfinished long-term goal in quantum information theory. As long as the multipartite system is concerned, the relation between the entanglement contained in different partitions or different subsystems need to take into account. The complete multipartite entanglement measure and the complete monogamy relation is a framework that just deals with such a issue. In this paper, we put forward conditions to justify whether the multipartite entanglement monotone (MEM) and genuine multipartite entanglement monotone (GMEM) are complete, completely monogamous, and tightly complete monogamous according to the feature of the reduced function. Especially, with the assumption that the maximal reduced function is nonincreasing on average under LOCC, we proposed a class of complete MEMs and a class of complete GMEMs via the maximal reduced function for the first time. By comparison, it is shown that, for the tripartite case, this class of GMEMs is better than the one defined from the minimal bipartite entanglement in literature under the framework of complete MEM and complete monogamy relation. In addition, the relation between monogamy, complete monogamy, and the tightly complete monogamy are revealed in light of different kinds of MEMs and GMEMs.  	
\end{abstract}


\maketitle
\end{CJK*}


\section{Introduction}

Entanglement, as one of the most puzzling features in quantum mechanics, has been widely used as an essential resource for quantum communication~\cite{Nielsen,Bennett1992communication,Zhang2006experimental}, quantum cryptography~\cite{Ekert1991prl,Giorgi2011prl}, and quantum computing~\cite{Ekert1998,Datta2005}, etc. 
The utility of an entangled state for these applications is often
directly related to the degree or type of entanglement contained in it. Therefore, efficiently quantifying and
characterizing multipartite entanglement is of paramount importance.
Especially, the genuine multipartite entanglement, as one of the important types of entanglement, offers significant advantages in quantum tasks compared with bipartite entanglement~\cite{Horodecki2009}.

The phenomenon becomes much more complex for multipartite entanglement, particularly the genuinely
multipartite entanglement, entanglement shared between all of the particles. Over the years, many multipartite entanglement measures have been proposed, such as the ``residual tangle'' which reports the genuine three-qubit
entanglement~\cite{Coffman}, the genuinely multipartite concurrence~\cite{Ma2011pra},
the $k$-ME concurrence~\cite{Hong2012pra}, the $m$ concurrence~\cite{Hiesmayr2008pra},  
the generalization of negativity~\cite{Jungnitsch2011prl}, the SL-invariant multipartite measure of entanglement~\cite{Verstraete2003pra,Luque2003pra,Osterloh2005pra,Osterloh2009jmp,Gour2010prl,Viehmann2011pra}, 
and the $\alpha$-entanglement entropy~\cite{Szalay}, concurrence triangle~\cite{Xie2021prl}, 
concentratable entanglement~\cite{Jacob2021prl}, geometric mean of bipartite concurrence~\cite{Liyinfei2022prr},
concurrence triangle induced genuine multipartite entanglement measure~\cite{Jin2022}, and a general way of constructing genuine multipartite entanglement monotone is proposed in Ref.~\cite{Guo2022jpa}.
In Ref.~\cite{Guo2020pra}, we proposed a framework of complete multipartite entanglement monotone from which the entanglement between any partitions or subsystems with the coarsening relation could be compared with each other.

In the context of describing multipartite entanglement, another fundamental task is to understand how entanglement is distributed over many parties since it reveals fundamental insights into the nature of quantum correlations~\cite{Horodecki2009} and has profound applications in both quantum communication~\cite{Terhal2004,Pawlowski} and other area of physics~\cite{streltsov2012are,Augusiak2014pra,Ma2011,Garcia,Lloyd}.
This characteristic trait of distribution is known as the monogamy law of entanglement~\cite{Osborne,Terhal2004}, which means that the more entangled two parties are, the less correlated they can be with other parties.
Quantitatively, the monogamy of entanglement is described by an inequality~\cite{Coffman,Osborne,streltsov2012are,
Dhar,Hehuan} or equality~\cite{GG,GG2019,Guo2020pra}, involving a bipartite entanglement monotone or multipartite entanglement monotone (MEM). Consequently, considerable research has been undertaken in this
direction~\cite{Koashi,Coffman,Osborne,streltsov2012are,Dhar,Hehuan,GG,GG2019,Guo2020pra}.

Very recently, we discussed when the genuine multipartite entanglement measure is complete~\cite{Guo2022entropy} with the same spirit as in Ref.~\cite{Guo2020pra}. Under such a sense, the hierarchy structure of the entanglement in the system is clear. Moreover, whether the multipartite entanglement measure is proper or not can be justified together with the framework of complete monogamy relation for the multipartite system established in Ref.~\cite{Guo2020pra}. The framework of complete monogamy relation is based on the complete multipartite entanglement measure~\cite{Guo2020pra,G2021qst,Guo2022entropy}. With this postulates, the distribution of entanglement appears more explicitly.

Multipartite entanglement measure is always defined via the bipartite entanglement measure. Let $\mS^X$ be the set of all density matrices acting on the state space $\mH^X$. Recall that, a function $E: \mS^{AB}\to\mbb{R}_{+}$ is called a 
measure of entanglement~\cite{Vedral97,Vedral98} if {(1)} $E(\sigma^{AB})=0$ for any separable density matrix $\sigma^{AB}\in\mS^{AB}$, and {(2)} $E$ behaves monotonically decreasing under local operations and classical communication (LOCC). Moreover, convex measures of entanglement that do not increase \emph{on average} under LOCC are called entanglement monotones~\cite{Vidal2000,Vedral98}. By replacing $\mS^{AB}$ with $\mS^{A_1A_2\cdots A_n}$, it is just the multipartite entanglement measure/monotone, and denoted by $E^{(n)}$. Any bipartite entanglement monotone corresponds to a concave function on the reduced state when it is evaluated for the pure states~\cite{Vidal2000}.
For any entanglement measure $E$, if 
\be\label{h}
h\left( \rho^A\right) = E\left( |\psi\lr\psi|^{AB}\right)
\ee 
is concave, i.e. $h[\lambda\rho_1+(1-\lambda)\rho_2]\geq\lambda h(\rho_1)+(1-\lambda)h(\rho_2)$
for any states $\rho_1$, $\rho_2$, and any $0\leq\lambda\leq1$,
then the convex roof extension of $E$, i.e., $E_F\left(\rho^{AB}\right)\equiv\min\sum_{j=1}^{n}p_jE\left(|\psi_j\lr\psi_j|^{AB}\right)$,
is an entanglement monotone, where the minimum is taken over all pure state decompositions of $\rho^{AB}=\sum_{j=1}^{n}p_j|\psi_j\lr\psi_j|^{AB}$. We call $h$ the \textit{reduced function} of $E$ and $\mH^A$ the \textit{reduced subsystem} throughout this paper. 

An $n$-partite pure state $|\psi\rangle\in \mathcal{H}^{A_1A_2\cdots A_n}$ is called biseparable if it can be written as
$|\psi\rangle=|\psi\rangle^X \otimes |\psi\rangle^Y$
for some bipartition of $A_1A_2\cdots A_n$ (for example, $A_1A_3|A_2A_4$ is a bipartition of $A_1A_2A_3A_4$). 
An $n$-partite mixed state $\rho$ is
biseparable if it can be written as a convex combination of
biseparable pure states
$\rho=\sum_{i}p_i|\psi_i\rangle \langle\psi_i|$, 
wherein the contained $\{|\psi_i\rangle\}$ can be biseparable with respect to different
bipartitions (i.e., a mixed biseparable state does not need to be separable with respect to any particular bipartition). If $\rho$ is not biseparable, then it is
called genuinely entangled.
A multipartite entanglement measure $E^{(n)}$ is called a genuine multipartite entanglement measure if (i) $E^{(n)}(\sigma)=0$ for any biseparable state $\sigma$, (ii) $E^{(n)}(\rho)>0$ for any genuine entangled state, and (iii) it is convex~\cite{Ma2011pra}. A genuine multipartite entanglement measure is called a genuine multipartite entanglement monotone (GMEM) if it does not increase on average under LOCC.

In Refs.~\cite{Guo2020pra,Guo2022jpa,Guo2022entropy}, we present MEMs and GMEMs that are defined by the sum of the reduced function on pure states and then extended to mixed states via the convex-roof structure. The aim of this paper is to give a condition that can justify when the MEMs and the GMEMs defined in this way is complete and completely monogamous. Moreover, we give another way of defining MEMs and the GMEMs from the maximal reduced function and then discuss when these quantities are complete and completely monogamous.

The remainder of this paper is organized as follows. In Sec. II, we introduce some preliminaries. Sec. III discusses the properties of the reduced functions of the entanglement monotones so far in literature. Sec. IV is divided into two subsections. Subsec. A discusses the MEM defined by the sum of reduced functions, and in Subsec. B, we give the MEMs defined by the maximal reduced function. Both of these two MEMs are explored under the framework of the complete measure and the complete monogamy relation. In Sec. V, we consider three kinds of GMEMs which are defined by the sum of reduced functions, the maximal reduced function, and the minimal reduced function, respectively, under the framework the complete measure and the complete monogamy relation. We present a conclusion in Sec. VI.

\section{Notations and Preliminaries}

The framework of the complete entanglement measure/monotone is closely related to the coarser relation of multipartite partition. We first introduce three kinds of coarser relation in Subsec. A, from which we then review the complete MEM, complete GMEM, monogamy relation and complete monogamy relation, respectively, in the latter three subsections.

\subsection{Coarser relation of multipartite partition}

Let $X_1|X_2| \cdots |X_{k}$ and $Y_1|Y_2| \cdots |Y_{l}$ be two partitions of $A_1A_2\cdots A_n$ or subsystem of $A_1A_2\cdots A_n$ (for instance, partition $AB|C|DE$ is a $3$-partition of the 5-particle system $ABCDE$ with $X_1=AB$, $X_2=C$ and $X_3=DE$).
We denote by~\cite{Guo2022entropy}
\bea\label{coarser}
X_1|X_2| \cdots| X_{k}\succ^a Y_1|Y_2| \cdots |Y_{l},\\
X_1|X_2| \cdots| X_{k}\succ^b Y_1|Y_2| \cdots |Y_{l},\\
X_1|X_2| \cdots| X_{k}\succ^c Y_1|Y_2| \cdots |Y_{l}~
\eea 
if $Y_1|Y_2| \cdots |Y_{l}$ can be obtained from $X_1|X_2| \cdots| X_{k}$
by 
\begin{itemize}
	\item[(a)] discarding some subsystem(s) of $X_1|X_2| \cdots| X_{k}$,
	\item[(b)] combining some subsystems of $X_1|X_2| \cdots| X_{k}$,
	\item[(c)] discarding some subsystem(s) of some subsystem(s) $X_t$ provided that $X_{t}=A_{t(1)}A_{t(2)}\cdots A_{t(f(t))}$ with $f(t)\geqslant2$, $1\leqslant t\leqslant k$,
\end{itemize}
respectively. For example, $A|B|C|D\succ^a A|B|D\succ^a B|D$,
$A|B|C|D\succ^b AC|B|D\succ^b AC|BD$, $A|BC\succ^c A|B$.
We call $Y_1|Y_2| \cdots |Y_{l}$ is coarser than $X_1|X_2| \cdots| X_{k}$ if 
 $Y_1|Y_2| \cdots |Y_{l}$ can be obtained from $X_1|X_2| \cdots| X_{k}$
by one or some of item (a)-item (c), and we denote it by $X_1|X_2| \cdots| X_{k}\succ Y_1|Y_2| \cdots |Y_{l}$.

Furthermore, if $X_1|X_2| \cdots| X_{k}\succ Y_1|Y_2| \cdots |Y_{l}$,
we denote by $\Xi(X_1|X_2| \cdots| X_{k}- Y_1|Y_2| \cdots |Y_{l})$ the set of
all the partitions that are coarser than $X_1|X_2| \cdots| X_{k}$ but (i) neither coarser than $Y_1|Y_2| \cdots |Y_{l}$ nor the one from which one can derive $Y_1|Y_2| \cdots |Y_{l}$ by the coarsening means, and (ii) either exclude any subsystem of $Y_1|Y_2| \cdots |Y_{l}$ or include both some subsystems of $Y_1|Y_2| \cdots |Y_{l}$ with all the subsystems $Y_j$s included are regarded as one subsystem and other subsystems in $X_1|X_2| \cdots| X_{k}$ but not in $Y_1|Y_2| \cdots| Y_{l}$, and (iii) if $Y_1|Y_2| \cdots |Y_{l}=X_1|X_2| \cdots|X_{l-1}|X_{l}\cdots X_{k}$, $\Xi(X_1|X_2| \cdots| X_{k}- Y_1|Y_2| \cdots |Y_{l})$ contains only $X_{l}|\cdots| X_{k}$ and the one coarser than it.
For example, $\Xi(A|B|CD|E-A|B)=\{ A|CD|E$, $A|CDE$, $ACD|E$, $AE|CD$, $A|C|E$, $A|D|E$,  
$AE|C$, $A|CE$,  $AC|E$, $A|DE$, $AE|D$, $AD|E$,  $A|CD$,  $CD|E$, 
$A|C$, $A|D$, $A|E$, $C|E$, $D|E$,  
$B|CD|E$, $B|CDE$, $BCD|E$, $BE|CD$, $B|D|E$, $B|C|E$,
$BE|C$,  $B|CE$, $BC|E$,  $B|DE$, $BE|D$, $BD|E$, $B|CD$,   
$B|C$, $B|D$, $B|E$, $AB|CD|E$, $AB|C|E$, $AB|D|E$, 
$AB|CDE$, $ABCD|E$, $ABE|CD$,
$AB|CD$, $AB|CE$, $AB|DE$, 
$ABE|C$, $ABE|D$, $ABC|E$, $ABD|E$, 
$AB|C$, $AB|D$,  $AB|E\}$,
$\Xi(A|B|C|D|E-A|B|C)$
$=\{A|D|E$, $AD|E$, $AE|D$, $A|DE$, $A|D$, $A|E$, $D|E$, 
$B|D|E$, $B|DE$, $BD|E$, $BE|D$, $B|D$, $B|E$,
$C|D|E$, $C|DE$, $CD|E$, $CE|D$, $C|D$, $C|E$, 
$AB|D|E$, $ABD|E$, $ABE|D$, $AB|DE$, $AB|D$, $AB|E$,
$AC|D|E$, $ACD|E$, $ACE|D$, $AC|DE$, $AC|D$, $AC|E$, 
$BC|D|E$, $BCD|E$, $BCE|D$, $BC|DE$, $BC|D$, $BC|E$, 
$ABC|D|E$, $ABC|DE$, $ABCD|E$, $ABCE|D$, $ABC|D$, $ABC|E\}$.
$\Xi(A|B|C|D-A|BCD)=\{B|C|D$, $B|CD$, $BC|D$, $C|BD$, $B|C$,
$C|D$, $B|D\}$.

\subsection{Complete MEM}

A multipartite entanglement measure $E^{(n)}$ is called a \textit{unified}
multipartite entanglement measure if it satisfies the unification condition~\cite{Guo2020pra}:
\begin{itemize}
	\item[(i)] (additivity): 
	\bea 
	E^{(n)}(A_1A_2\cdots A_k\ot{A_{k+1}\cdots A_n})\quad\quad\quad\quad\quad\quad\nonumber
	\\
	=E^{(k)}({A_1A_2\cdots A_k})+E^{(n-k)}({A_{k+1}\cdots A_n}),
	\eea 
	holds for all $\rho^{A_1A_2\cdots A_n}\in\mS^{A_1A_2\cdots A_n}$, hereafter $E^{(n)}(X)$ refers to $E^{(n)}(\rho^X)$ and $E^{(1)}\equiv0$;
	\item[(ii)] (permutation invariance): $E^{(n)}({A_1A_2\cdots A_n})=E^{(n)}({A_{\pi(1)}A_{\pi(2)}\cdots A_{\pi(n)}})$,
	for all $\rho^{A_1A_2\cdots A_n}\in\mS^{A_1A_2\cdots A_n}$ and any permutation $\pi$;
	\item[(iii)] (coarsening monotone): 
	\bea\label{coarsen}
	E^{(k)}(X_1|X_2| \cdots| X_{k})\geqslant E^{(l)}(Y_1|Y_2| \cdots |Y_{l})
	\eea
	holds for all $\rho^{A_1A_2\cdots A_n}\in\mS^{A_1A_2\cdots A_n}$ whenever $X_1|X_2| \cdots| X_{k}\succ^a Y_1|Y_2| \cdots |Y_{l}$,
	where $X_1|X_2| \cdots |X_{k}$ and $Y_1|Y_2| \cdots |Y_{l}$ are two partitions of $A_1A_2\cdots A_n$ or subsystem of $A_1A_2\cdots A_n$, the vertical bar indicates the split 
	across which the entanglement is measured.
\end{itemize}
$E^{(n)}$ is called a \textit{complete}
multipartite entanglement measure if it satisfies both the conditions above and the hierarchy condition~\cite{Guo2020pra}: 
\begin{itemize}
\item[(iv)] (tight coarsening monotone): Eq.~\eqref{coarsen} holds for all $\rho\in\mS^{A_1A_2\cdots A_n}$ whenever $X_1|X_2|\cdots|X_{k}\succ^b Y_1|Y_2|\cdots |Y_{l}$.
\end{itemize}

\subsection{Complete GMEM}

Let $E_g^{(n)}$ be a genuine multipartite entanglement measure. It is defined to be a \textit{unified} genuine multipartite entanglement measure if it satisfies the \textit{unification condition}~\cite{Guo2022entropy}, i.e., 
\begin{itemize}
	\item[(i)] (permutation invariance): $E_g^{(n)}({A_1A_2\cdots A_n})=E_g^{(n)}({A_{\pi(1)}A_{\pi(2)}\cdots A_{\pi(n)}})$,
	for all $\rho^{A_1A_2\cdots A_n}\in\mS^{A_1A_2\cdots A_n}$ and any permutation $\pi$;
	\item[(ii)] (coarsening monotone): 
	\bea\label{gcoarsen}
	E_g^{(k)}(X_1|X_2|\cdots|X_{k})> E_g^{(l)}(Y_1|Y_2|\cdots|Y_{l})
	\eea
	holds for all $\rho^{A_1A_2\cdots A_n}\in\mS_g^{A_1A_2\cdots A_n}$ whenever $X_1|X_2| \cdots| X_{k}\succ^a Y_1|Y_2| \cdots |Y_{l}$.
\end{itemize}
A unified GMEM $E_g^{(n)}$ is call a \textit{complete} genuine multipartite entanglement measure 
if $E^{(n)}_{g}$ admits the \textit{hierarchy condition}~\cite{Guo2022entropy}, i.e.,
\begin{itemize}
	\item[(iii)] (tight coarsening monotone): 
	\bea\label{ghierarchy}
	E_g^{(k)}(X_1|X_2|\cdots|X_{k})\geq E_g^{(l)}(Y_1|Y_2| \cdots |Y_{l})
	\eea 
	holds for all $\rho\in\mS_g^{A_1A_2\cdots A_n}$ whenever $X_1|X_2|\cdots|X_{k}\succ^b Y_1|Y_2|\cdots |Y_{l}$.
\end{itemize}

\subsection{Monogamy Relation}

For an bipartite entanglement measure $E$, $E$ is said to be monogamous if~\cite{Coffman,Koashi}
\bea\label{monogamy1}
E(A|BC)\geqslant E(AB)+E(AC).
\eea
However, Equation~(\ref{monogamy1}) is not valid for many entanglement measures~\cite{Coffman,Dhar,GG} but some power function of $Q$ admits the monogamy relation (i.e., $E^\alpha(A|BC)\geqslant E^\alpha(AB)+E^\alpha(AC)$ for some $\alpha>0$). In Ref.~\cite{GG}, we improved the definition of monogamy as: A bipartite measure of entanglement $E$ is monogamous if for any $\rho\in\mS^{ABC}$ that satisfies the \textit{disentangling condition},~i.e.,    
\be\label{cond1}
E( \rho^{A|BC}) =E( \rho^{AB}),
\ee
we have that $E(\rho^{AC})=0$, where $\rho^{AB}=\tr_C\rho^{ABC}$. With respect to this definition, a continuous measure $E$ is monogamous according to this definition if and only if there exists $0<\alpha<\infty$ such that
\be\label{power}
E^\alpha( \rho^{A|BC}) \geqslant E^\alpha( \rho^{AB}) +E^\alpha( \rho^{AC})	
\ee
for all $\rho$ acting on the state space $\mH^{ABC}$ with fixed $\dim\mH^{ABC}=d<\infty$ (see Theorem 1 in Ref.~\cite{GG}).

In Ref.~\cite{Guo2020pra}, in order to characterize the distribution of entanglement in a ``complete'' sense, the term ``complete monogamy'' of the unified multipartite entanglement measure is proposed. For a unified multipartite entanglement measure $E^{(n)}$, it is said to be {\textit{completely monogamous}} if for any
$\rho\in\mathcal{S}^{A_1A_2\cdots A_n}$ that satisfies~\cite{Guo2020pra}
\bea\label{cond2}
E^{(k)}(X_1|X_2| \cdots| X_{k})= E^{(l)}(Y_1|Y_2| \cdots |Y_{l})
\eea
with $X_1|X_2| \cdots| X_{k}\succ^a Y_1|Y_2| \cdots |Y_{l}$ we have that
\bea\label{cond2x}
E^{(\ast)}({\Gamma}) =0
\eea
holds for all $\Gamma\in \Xi(X_1|X_2| \cdots| X_{k}- Y_1|Y_2| \cdots |Y_{l})$, hereafter the superscript $(\ast)$ is associated with the partition $\Gamma$, e.g., if $\Gamma$ is a $n$-partite partition, then $(\ast)=(n)$. 
For example, $E^{(3)}$ is completely monogamous if for any $\rho^{ABC}$ that admits $E^{(3)}(ABC)=E^{(2)}(AB)$ we get $E^{(2)}(AC)=E^{(2)}(BC)=0$. Let $E^{(n)}$ be a complete multipartite entanglement measure. $E^{(n)}$ is defined to be
\textit{tightly complete monogamous} if for any $\rho\in\mathcal{S}^{A_1A_2\cdots A_n}$ that satisfies~\cite{Guo2020pra}
\bea\label{cond3}
E^{(k)}(X_1|X_2| \cdots| X_{k})= E^{(l)}(Y_1|Y_2| \cdots |Y_{l})
\eea
with $X_1|X_2| \cdots| X_{k}\succ^b Y_1|Y_2| \cdots |Y_{l}$ we have that
\bea\label{cond3x}
E^{(\ast)}({\Gamma}) =0
\eea
holds for all $\Gamma\in \Xi(X_1|X_2| \cdots| X_{k}- Y_1|Y_2| \cdots |Y_{l})$. For instance, $E^{(3)}$ is tightly complete monogamous if for any $\rho^{ABC}$ that admits $E^{(3)}(ABC)=E^{(2)}(A|BC)$ we have $E^{(2)}(BC)=0$.

Let $E_g^{(n)}$ be a genuine multipartite entanglement measure. We denote by $S_g^{A_1A_2\cdots A_m}$ the set of all genuine entangled states in $S^{A_1A_2\cdots A_m}$. $E_g^{(n)}$ is completely monogamous if it obeys Eq.~\eqref{gcoarsen}~\cite{Guo2022entropy}.	A complete genuine multipartite entanglement measure $E_g^{(n)}$ is
tightly complete monogamous if it satisfies the \textit{genuine disentangling condition}, i.e., either for any $\rho\in\mS_g^{A_1A_2\cdots A_m}$ that satisfies~\cite{Guo2022entropy}  
\bea\label{g-tight}
E_g^{(k)}({X_1|X_2|\cdots|X_{k}})=E_g^{(l)}({Y_1|Y_2|\cdots|Y_{l}}) 
\eea
with $X_1|X_2| \cdots| X_{k}\succ^b Y_1|Y_2| \cdots |Y_{l}$ we have that
\bea\label{g-tight1}
E_g^{(\ast)}({\Gamma})=0
\eea
holds for all $\Gamma\in \Xi(X_1|X_2| \cdots| X_{k}- Y_1|Y_2| \cdots |Y_{l})$, or 
\bea\label{g-tight2}
E_g^{(k)}({X_1|X_2|\cdots|X_{k}})>E_g^{(l)}({Y_1|Y_2|\cdots|Y_{l}}) 
\eea
holds for any $\rho\in\mS_g^{A_1A_2\cdots A_m}$.

In Ref.~\cite{Guo2020pra}, we showed that the tightly complete monogamy is stronger than the complete monogamy for the
complete MEMs that defined by the convex-roof extension. One can easily find that it is also true for any complete GMEM defined by the convex-roof extension.

\section{Strict concavity and subadditivity of the reduced function}

Any entanglement monotone, when evaluated on pure states, is uniquely determined by its reduced function and vice versa. Therefore, the feature of the entanglement monotone defined via the convex-roof extension rests with the quality of its reduced function. In Ref.~\cite{GG2019}, we proved that the bipartite entanglement monotone is monogamous whenever its reduced function is strictly concave. In ths Section, we review all the reduced functions of the entanglement monotones in literature so far and then discuss the subadditivity of these functions. As what we will show in the next two Sections, the subadditivity is affinitive with the completeness of the measures for some kind of MEM/GMEM.

\subsection{Strict concavity}

The reduced functions of the entanglement of formation $E_f$~\cite{Bennett1996prl,Horodecki01}, tangle $\tau$~\cite{Rungta2003pra}, concurrence $C$~\cite{Hill,Wootters,Rungta}, negativity $N$~\cite{Lee},
the Tsallis $q$-entropy of entanglement $E_q$~\cite{Kim2010pra}, and the R\'{e}nyi $\alpha$-entropy of entanglement $E_\alpha$~\cite{Vidal2000,Kim2010jpa} are
\beax
h(\rho)&=&S(\rho),\\
h_{\tau}(\rho)&=&h_C^2(\rho)=2(1-\tr\rho^2),\\
h_N(\rho)&=&\frac12[(\tr\sqrt{\rho})^2-1],\\
h_q(\rho)&=&\dfrac{1-\tr\rho^q}{q-1},\quad q>0,\\
h_\alpha(\rho)&=&(1-\alpha)^{-1}\ln(\tr\rho^\alpha),\quad 0<\alpha<1,
\eeax
respectively, where $S$ is the von Neumann entropy. It has been shown that $h$, $h_{\tau}$, $h_C$, $h_N$, $h_q$, and $h_\alpha$ are not only concave but also strictly concave~\cite{Wehrl1978,Vidal2000,GG2019} (where the strict concavity of $h_N$ is proved very recently in Ref.~\cite{Guo2022prxsub}).

The reduced functions of the entanglement monotones induced by the fidelity-based distances $E_{\mathcal{F}}$,  $E_{\mathcal{F}'}$, and $E_{A\mathcal{F}}$ are~\cite{Guo2020qip}
\beax
h_{\mathcal{F}}(\rho)&=&1-\tr\rho^3,\\
h_{\mathcal{F}'}(\rho)&=&1-\left( \tr\rho^2\right)^2 ,\\
h_{A\mathcal{F}}(\rho)&=&1-\sqrt{\tr\rho^3},
\eeax
respectively. They are strictly concave~\cite{Guo2022entropy}.

In Ref.~\cite{Guo2022prxsub}, four kinds of partial norm of entanglement are investigated: the partial-norm of entanglement $E_2$, the minimal partial-norm of entanglement $E_{\min}$, the reinforced minimal partial-norm of entanglement $E_{\min'}$, and the partial negativity $\hat{N}$. The reduced functions of $E_2$, $E_{\min}$, $E'_{\min}$, and $\hat{N}$ are
\beax
h_2(\rho)&=&1-\|\rho\|,\\
h_{\min}(\rho)&=&\|\rho\|_{\min},\\
h_{\min'}(\rho)&=&r(\rho)\|\rho\|_{\min},\\
\hat{h}(\rho)&=&\sqrt{\delta_1\delta_2},
\eeax
where $r(\rho)$ denotes the rank of $\rho$, $\|\cdot\|$ is the operator norm, i.e., $\|X\|=\sup_{|\psi\ra}\|A|\psi\ra\|$,
\beax
\|\rho\|_{\min}=\begin{cases}~\lambda_{\min}^2,& \lambda_{\min}<1,\\
	~0, &\lambda_{\min}=1,
\end{cases}
\eeax 
and $\delta_1$, $\delta_2$ are the two largest eigenvalues of $\rho$. All of them are concave but not strictly concave ($\hat{h}$ is only strictly concave on qubit states), and these entanglement monotones are not monogamous~\cite{Guo2022prxsub}.

\subsection{Subadditivity}

We summarize the subadditivity of the reduced functions in literature as following: 
\begin{itemize}
	\item[(i)] $S$ is additive and subadditive~\cite{Wehrl1978}, i.e.,
	\bea
	S(\rho\ot\sigma)=S(\rho)+S(\sigma)
	\eea 
	and 
	\bea 
	S(\rho^{AB})\leq S(\rho^A)+S(\rho^B),
	\eea 
	respectively.
	
	\item[(ii)] $S_q$ is subadditive iff $q>1$, but not additive, and
	for $0<q<1$, $S_q$ is neither subadditive nor superadditive~\cite{Raggio}
	(superadditivity refers to $S_q(\rho^{AB})\geqslant S_q(\rho^A)+S_q(\rho^B)$).
	In addition, 
	\bea
	S_q(\rho^{A}\ot\rho^B)= S_q(\rho^A)+S_q(\rho^B)
	\eea
	iff $\rho^A$ or $\rho^B$ is pure~\cite{Raggio}.
	\item[(iii)] $h_\alpha$ is additive but not subadditive~\cite{Beck,Aczel}.
	
	\item[(iv)] $h_\tau$ is subadditive~\cite{Audenaerta2007jmp}, i.e.,
	\bea
	1+\tr\rho_{AB}^2\geq \tr\rho_A^2+\tr\rho_B^2.
	\eea
	In particular, the equality holds iff $\rho^A$ or $\rho^B$ is pure~\cite{Guo2020pra}.
	
	\item[(v)] $h_N$ is neither subadditive nor supperadditive~\cite{Guo2020pra}.
\end{itemize}
Item (iv) implies $h_C$ is subadditive and the equality holds iff $\rho^A$ or $\rho^B$ is pure. $h_{\mathcal{F}}$ is subadditive since it coincides with $S_q/2$ ($q=3$). We conjecture that $h_{\mathcal{F}'}$ and $h_{A\mathcal{F}}$ are subadditive.

\begin{pro}\label{subadditive-h2}
	$h_2$ is subadditive, i.e., 
	\bea\label{subadditive}
	1+\|\rho^{AB}\|\geqslant\|\rho^A\|+\|\rho^B\|
	\eea
	holds for any $\rho^{AB}\in\mS^{AB}$. In particular, the equality holds iff $\rho^A$ or $\rho^B$ is a pure state.
\end{pro}

\begin{proof}
	Note that partial trace is a quantum channel and any quantum channel can be regarded as a operator on the space of the trace-class operators. The norm of quantum channel in such a sense is 1. Therefore $\|\rho^{AB}\|\geqslant\|\rho^{A,B}\|$. Moreover, if $1+\|\rho^{AB}\|=\|\rho^A\|+\|\rho^B\|$, then $\|\rho^A\|=1$ or  $\|\rho^B\|=1$, which completes the proof.
\end{proof}

\begin{table}
	\caption{\label{tab:table1} Comparing of the properties of the reduced functions.
		C, SC, SA, and A signify the function is concave, strictly concave, subadditive, and additive, respectively.}	
	\begin{ruledtabular}
		\begin{tabular}{cccccc}
			$E$  &$h$&     C        & SC   &  SA & A  \\ \colrule
			$E_f$  &$S$                          &$\checkmark$         &$\checkmark$       &$\checkmark$&$\checkmark$\\
			$C$    &$\sqrt{2(1-\tr\rho^2)}$      &$\checkmark$         &$\checkmark$       &$\checkmark$&$\times$\\
			$\tau$ &$2(1-\tr\rho^2)$             &$\checkmark$         &$\checkmark$       &$\checkmark$&$\times$  \\
			$E_q$  &$\frac{1-\tr\rho^q}{q-1}$    &$\checkmark$($q>0$)  &$\checkmark$($q>1$)&$\checkmark$($q>1$) &$\times$\\
			$E_\alpha$&$\frac{\ln(\tr\rho^\alpha)}{1-\alpha}$, $\alpha\in(0,1)$
			&$\checkmark$       &$\checkmark$       &$\times$    &$\checkmark$ \\
			$N_F$  &$\frac{(\tr\sqrt{\rho})^2-1}{2}$   &$\checkmark$         &$\checkmark$       &$\times$    &$\times$\\
			$E_{\mF}$&$1-\tr\rho^3$                &$\checkmark$         &$\checkmark$       &$\checkmark$&$\times$\\
			$E_{\mF'}$&$1-(\tr\rho^2)^2$            &$\checkmark$         &$\checkmark$       &$\checkmark\footnotemark[1]$           &$\times$\\
			$E_{A\mF}$&$1-\sqrt{\tr\rho^3}$         &$\checkmark$         &$\checkmark$       &$\checkmark\footnotemark[1]$           &$\times$\\
			$E_2$	   &$1-\|\rho\|$                 &$\checkmark$         &$\times$           &$\checkmark$&$\times$\\
			$E_{\min}$&$\|\rho\|_{\min}$            &$\checkmark$         &$\times$           &$\times$           &$\times$\\
			$E_{\min'}$&$r(\rho)\|\rho\|_{\min}$     &$\checkmark$         &$\times$           &$\times$            &$\times$\\
			$\hat{N}$  &$\sqrt{\delta_1\delta_2}$    &$\checkmark\footnotemark[2]$         &$\times$           &$\checkmark\footnotemark[1]$            &$\times$
		\end{tabular}
	\end{ruledtabular}
	\footnotetext[1]{We conjecture that they are subadditive.}
	\footnotetext[2]{We conjecture that it is concave.}
\end{table}

Let 
\beax
\rho^{AB}=\frac12|\psi\ra\la\psi|+\frac12|\phi\ra\la\phi|
\eeax
with $|\psi\ra=\sqrt{\frac45}|00\ra+\sqrt{\frac15}|11\ra$ and $|\phi\ra=\sqrt{\frac45}|22\ra+\sqrt{\frac15}|33\ra$.
It is clear that $\|\rho^{AB}\|_{\min}=\frac12>\|\rho^A\|_{\min}+\|\rho^B\|_{\min}=1/10+1/10=1/5$.
That is, $\|\cdot\|_{\min}$ is not subadditive. Clearly, $h_{\min'}$ is also not subadditive.
According to Proposition~\ref{subadditive-h2}, $h_{\min}$ and $h_{\min'}$ are subadditive on the states that satisfies $r(\rho^{AB})=r(\rho^{A})=r(\rho^{B})=2$. One can easily verifies that $h_2$, $h_{\min}$, $h_{\min'}$, and $\hat{h}$ are not additive.

We conjecture that $\hat{h}$ is subadditive, i.e., 
\bea
\hat{h}(\rho^{AB})\leqslant\hat{h}(\rho^{A})+\hat{h}(\rho^{B})
\eea
holds for any $\rho^{AB}\in\mS^{AB}$. In what follows, we always assume that $h_{\mathcal{F}'}$, $h_{A\mathcal{F}}$, and $\hat{h}$ are subadditive, and that $\hat{h}$ is concave.

The reduced functions of parametrized entanglement monotones in Ref.~\cite{Yangxue2021pra} and Ref.~\cite{Wei2022jpa} are
\beax
h_{q'}(\rho)=1-\tr\rho^q, \quad q>1, 
\eeax
and
\beax h_{\alpha'}(\rho)=\tr\rho^\alpha-1,\quad  0<\alpha<1,
\eeax 
respectively. Obviously, the properties of these two functions above are the same as that of $h_q$, although they are different from $E_q$~\cite{Yangxue2021pra,Wei2022jpa}. We summarize the properties of these reduced functions in Table~\ref{tab:table1} for more convenience.

\section{Complete MEM}

\subsection{Complete MEM from sum of the reduced functions}

In Ref.~\cite{Guo2020pra}, we put forward several complete MEMs defined by the sum of the reduced functions on all the single subsystems. In fact, this scenario is valid for all entanglement monotones. Let $|\psi\ra^{A_1A_2\cdots A_n}$ be a pure state in $\mH^{A_1A_2\cdots A_n}$ and $h$ be a non-negative concave function on $\mS^X$. We define
\bea\label{sum1}
E^{(n)}(|\psi\ra^{A_1A_2\cdots A_n})=\frac{1}{2}\sum_ih(\rho^{A_i})
\eea
and then extend it to mixed states by the convex-roof structure. We denote $E^{(n)}$ by $E_f^{(n)}$, $C^{(n)}$, $\tau^{(n)}$, $E_q^{(n)}$, $E_\alpha^{(n)}$, $N_F^{(n)}$, $E_{\mathcal{F}}^{(n)}$, $E_{\mathcal{F}'}^{(n)}$, $E_{A\mathcal{F}}^{(n)}$, $E_2^{(n)}$, $E_{\min}^{(n)}$, ${E}_{\min'}^{(n)}$, and $\hat{N}^{(n)}$ whenever
$h=S$, $h_C$, $h_\tau$, $h_q$, $h_\alpha$, $h_N$, $h_{\mathcal{F}}$, $h_{\mathcal{F}'}$, $h_{A\mathcal{F}}$, $h_2$, $h_{\min}$, $h_{\min'}$, and $\hat{h}$, respectively. Here, $E_f^{(n)}$, $C^{(n)}$, $\tau^{(n)}$, $E_q^{(n)}$, $E_\alpha^{(n)}$, and $N_F^{(n)}$ have been discussed in Ref.~\cite{Guo2020pra} for the first time. The coefficient ``1/2'' is fixed by the unification condition when $E^{(n)}$ is regarded as a unified MEM. One need note here that $E_{\mathcal{F}}^{(n)}$, $E_{\mathcal{F}'}^{(n)}$, and $E_{A\mathcal{F}}^{(n)}$ are different from $E_{\mathcal{F},F}^{(n)}$, $E_{\mathcal{F}',F}^{(n)}$, and $E_{A\mathcal{F},F}^{(n)}$ respectively in Ref.~\cite{Guo2020qip}.

\begin{theorem}\label{th1}
Let $E^{(n)}$ be a non-negative function defined as in Eq.~\eqref{sum1}. Then the following statements hold true.
\begin{itemize}
	\item[(i)] $E^{(n)}$ is a unified MEM and is completely monogamous;
	\item[(ii)] $E^{(n)}$ is a complete MEM iff $h$ is subadditive;
	\item[(iii)] $E^{(n)}$ is tightly complete monogamous iff $h$ is subadditive with 
	\bea\label{sa-h}
	h(\rho^{AB})=h(\rho^A)+h(\rho^B)\Rightarrow \rho^{AB}~\mbox{is separable.}
    \eea
\end{itemize}
\end{theorem}

\begin{proof}
	We only need to discuss the case of $n=3$ with no loss of generality.
	
	(i) For any $|\psi\ra^{ABC}\in\mH^{ABC}$, we let 
	$E^{(2)}(\rho^{AB})=\sum_ip_iE^{(2)}(|\psi_i\ra)=\frac{1}{2}\sum_ip_i[h(\rho_i^A)+h(\rho_i^B)]$. 
	Then 
	\beax
	&&E^{(3)}(|\psi\ra^{ABC})=\frac{1}{2}\left[ h(\rho^A)+h(\rho^B)+h(\rho^C)\right] \\
	&\geq&\frac{1}{2}\left[ h(\rho^A)+h(\rho^B)\right]
	\geq\frac{1}{2}\sum_ip_i[h(\rho_i^A)+h(\rho_i^B)]\\
	&=&E^{(2)}(\rho^{AB}).
	\eeax
	That is, $E^{(3)}$ satisfies Eq.~\eqref{coarsen} for pure states and it is completely monogamous on pure states.
	For any mixed state $\rho^{ABC}$, we let $E^{(3)}(\rho^{ABC})=\sum_jq_jE^{(3)}(|\psi_j\ra)$ and $E^{(2)}(\rho_j^{AB})=\sum_ip_{i(j)}E^{(2)}(|\psi_{i(j)}\ra)=\frac{1}{2}\sum_ip_{i(j)}[h(\rho_{i(j)}^A)+h(\rho_{i(j)}^B)]$.
	Then 
	\beax
	&&E^{(3)}(\rho^{ABC})=\frac{1}{2}\sum_jq_j\left[ h(\rho_j^A)+h(\rho_j^B)+h(\rho_j^C)\right] \\
	&\geq&\frac{1}{2}\sum_jq_j\left[h_j(\rho^A)+h_j(\rho^B)\right]\\
	&\geq&\frac{1}{2}\sum_{i,j}q_jp_{i(j)}[h(\rho_{i(j)}^A)+h(\rho_{i(j)}^B)]
	\geq E^{(2)}(\rho^{AB}),
	\eeax
	i.e., it is a unified MEM. If $E^{(3)}(\rho^{ABC})=E^{(2)}(\rho^{AB})$, it yields $h(\rho_j^C)=0$ for any $j$, and
	thus $|\psi_j\ra^{ABC}=|\psi_j\ra^{AB}|\psi_j\ra^C$. Therefore it is completely monogamous. 
	
	(ii) If $E^{(3)}$ is a complete MEM, then $E^{(3)}(|\psi\ra^{ABC})\geq E^{(2)}(|\psi\ra^{A|BC})$ for any $|\psi\ra^{ABC}$, which implies $h(\rho^{BC})\leq h(\rho^B)+h(\rho^C)$. That is, $h$ is subadditive since $|\psi\ra^{ABC}$ is arbitrarily given. Conversely, if $h$ is subadditive, then $E^{(3)}(|\psi\ra^{ABC})\geq E^{(2)}(|\psi\ra^{A|BC})$ for any pure state $|\psi\ra^{ABC}$. For any mixed state $\rho^{ABC}$, we let
	$E^{(3)}(\rho^{ABC})=\sum_jq_jE^{(3)}(|\psi_j\ra)$.
	Then 
	\beax
	&&E^{(3)}(\rho^{ABC})=\frac{1}{2}\sum_jq_j\left[ h(\rho_j^A)+h(\rho_j^B)+h(\rho_j^C)\right] \\
	&\geq&\frac{1}{2}\sum_jq_j\left[h_j(\rho^A)+h_j(\rho^{BC})\right]
	\geq E^{(2)}(\rho^{A|BC}),
	\eeax
	i.e., it is a complete MEM. 
	
	(iii) It can be easily checked using the argument analogous to that of (ii) together with the fact that, if $E^{(n)}$ is tightly complete monogamous, it is automatically a complete MEM. 
\end{proof}

By Theorem~\ref{th1}, we can conclude:
(i) $E_f^{(n)}$, $C^{(n)}$, $\tau^{(n)}$, $E_q^{(n)}$, $E_\alpha^{(n)}$, $N_F^{(n)}$,
$E_{\mathcal{F}}^{(n)}$, $E_{\mathcal{F}'}^{(n)}$, $E_{A\mathcal{F}}^{(n)}$, 
$E_2^{(n)}$, $E_{\min}^{(n)}$, ${E}_{\min'}^{(n)}$, and $\hat{N}^{(n)}$ are unified MEMs and are completely monogamous;
(ii) $E_f^{(n)}$, $C^{(n)}$, $\tau^{(n)}$, $E_q^{(n)}$, 
$E_{\mathcal{F}}^{(n)}$, $E_{\mathcal{F}'}^{(n)}$, $E_{A\mathcal{F}}^{(n)}$, 
$E_2^{(n)}$, and $\hat{N}^{(n)}$ are complete MEMs;
(iii) $E_\alpha^{(n)}$, $N_F^{(n)}$, $E_{\min}^{(n)}$, and ${E}_{\min'}^{(n)}$ are not complete MEMs since the associated reduced functions are not subadditive which violate the hierarchy condition for some states.
(iv) $E_f^{(n)}$, $C^{(n)}$, $\tau^{(n)}$, and $E_2^{(n)}$ are tightly complete monogamous. However $E_2^{(n)}$, $E_{\min}^{(n)}$, ${E}_{\min'}^{(n)}$, and $\hat{N}^{(n)}$ are not monogamous.Together with Theorem in Ref.~\cite{GG2019}, we obtain that, for these MEMs, both monogamy and tightly complete monogamy are stronger than the complete monogamy under the frame work of the complete MEM, and that monogamy is stronger than both complete monogamy and tightly complete monogamy (e.g., $E_2^{(n)}$).

In particular, if $h$ is subadditive with $h(\rho^{AB})=h(\rho^A)+h(\rho^B)$ implies $\rho^{AB}=\rho^A\ot\rho^B$,
then $E^{(n)}$ is tightly complete monogamous. $S$, $h_\tau$, $h_C$, and $h_2$ belong to such situations.
We also conjecture that $h_q$, $h_\mF$, $h_{\mF'}$, $h_{A\mF}$, and $\hat{h}$ belong to such situations as well.
That is, we conjecture that $E_q^{(n)}$, $E_{\mathcal{F}}^{(n)}$, $E_{\mathcal{F}'}^{(n)}$, $E_{A\mathcal{F}}^{(n)}$, 
and $\hat{N}^{(n)}$ are tightly complete monogamous.

\begin{table}
	\caption{\label{tab:table2} Comparing of $E^{(n)}$ with different different reduced functions, and $\mE^{(n)}$.
		CM and TCM signify the measure is completely monogamous and tightly complete monogamous, respectively.}	
	\begin{ruledtabular}
		\begin{tabular}{ccccc}
			MEM     & Unified & Complete & CM & TCM \\ \colrule
			$E_f^{(n)}$         &$\checkmark$&$\checkmark$ &$\checkmark$&$\checkmark$\\
			$C^{(n)}$           &$\checkmark$&$\checkmark$ &$\checkmark$&$\checkmark$\\
			$\tau^{(n)}$        &$\checkmark$&$\checkmark$ &$\checkmark$&$\checkmark$  \\
			$E_q^{(n)}$         &$\checkmark$&$\checkmark$ &$\checkmark$&$\checkmark\footnotemark[1]$\\
			$E_\alpha^{(n)}$    &$\checkmark$&$\times$     &$\checkmark$&$\times$   \\
			$N_F^{(n)}$         &$\checkmark$&$\times$     &$\checkmark$&$\times$\\
			$E_{\mF}^{(n)}$     &$\checkmark$&$\checkmark$ &$\checkmark$&$\checkmark\footnotemark[1]$\\
			$E_{\mF'}^{(n)}$    &$\checkmark$&$\checkmark\footnotemark[2]$&$\checkmark$&$\checkmark\footnotemark[1]$\\
			$E_{A\mF}^{(n)}$    &$\checkmark$&$\checkmark\footnotemark[2]$&$\checkmark$&$\checkmark\footnotemark[1]$\\
			$E_{2}^{(n)}$       &$\checkmark$&$\checkmark$ &$\checkmark$&$\checkmark$\\
			$E_{\min}^{(n)}$    &$\checkmark$&$\times$     &$\checkmark$&$\times$\\
			${E}_{\min'}^{(n)}$ &$\checkmark$&$\times$     &$\checkmark$&$\times$\\
			$\hat{N}^{(n)}$     &$\checkmark$&$\checkmark\footnotemark[2]$&$\checkmark$&$\checkmark\footnotemark[1]$\\   \colrule
			${\mE}^{(n)}~(n\geq4)$          &$\checkmark$&$\checkmark$ &$\checkmark$&$\checkmark$
		\end{tabular}
	\end{ruledtabular}
	\footnotetext[1]{It is tightly complete monogamous under the assumption that $h$ is subadditive and Eq.~\eqref{sa-h} holds.}
	\footnotetext[2]{It is complete under the assumption that $h$ is subadditive.}
\end{table}

In Ref.~\cite{Guo2022jpa}, we put forward several multipartite entanglement measures which are defined by the sum of all bipartite entanglement. Let $|\psi\ra^{A_1A_2\cdots A_n}$ be a pure state in $\mH^{A_1A_2\cdots A_n}$ and $h$ be a non-negative concave function on $\mS^X$. We define~\cite{Guo2022jpa}
\bea\label{sum2}
\mE^{(n)}(|\psi\ra^{A_1A_2\cdots A_n})\quad\quad\quad\quad\quad\quad\quad\quad\quad\quad\quad\quad\quad\quad\quad~~\quad\nonumber\\
=\begin{cases}
	\frac{1}{2}\sum\limits_{i_1\leq \cdots\leq i_s, s< n/2}h(\rho^{A_{i_1}A_{i_2}\cdots A_{i_s}}),& \mbox{if}~n~\mbox{is odd},\\
	\frac{1}{2}\sum\limits_{i_1\leq \cdots\leq i_s<n, s\leq n/2}h(\rho^{A_{i_1}A_{i_2}\cdots A_{i_s}}),& \mbox{if}~n~\mbox{is even},
\end{cases}\quad~
\eea
for pure states and for mixed states by the convex-roof structure. Note that $\mE^{(n)}$ is just $\mE_{12\cdots n(2)}$ in Ref.~\cite{Guo2022jpa} provided that the corresponding bipartite entanglement measure is an entanglement monotone.
Clearly, 
\bea
E^{(n)}\leq\mE^{(n)},
\eea
and in general, $E^{(n)}<\mE^{(n)}$ whenever $n\geq4$. Indeed, we can easily show that $E^{(n)}(|\psi\ra)<\mE^{(n)}(|\psi\ra)$ iff $|\psi\ra$ is not fully separable, i.e., $|\psi\ra\neq|\psi\ra^{A_1}|\psi\ra^{A_2}\ot\cdots\ot|\psi\ra^{A_n}$. $\mE^{(3)}$ coincides with $E^{(3)}$ but $\mE^{(n)}$ is different from $E^{(n)}$ whenever $n\geq4$. The following Proposition is straightforward by the definition of $\mE^{(n)}$.

\begin{pro}\label{th12}
	Let $\mE^{(n)}$ be a non-negative function defined as in Eq.~\eqref{sum2}, $n\geq4$. Then $\mE^{(n)}$ is a complete MEM and it is completely monogamous and tightly complete monogamous.
\end{pro}

Then, when $n\geq4$, all these MEMs $\mE^{(n)}$ with the reduced functions we mentioned above are complete MEMs, and are not only completely monogamous but also tightly complete monogamous. We compare all these MEMs in Table~\ref{tab:table2} for convenience.

\subsection{Complete MEM from the maximal reduced function}

Let $|\psi\ra^{A_1A_2\cdots A_n}$ be a pure state in $\mH^{A_1A_2\cdots A_n}$ and $h$ be a non-negative concave function. We define
\bea\label{max1}
{E'}^{(n)}(|\psi\ra^{A_1A_2\cdots A_n})
=\max\limits_ih(\rho^{A_i})
\eea
and then extend it to mixed states by the convex-roof structure. By definition, ${E'}^{(n)}\leq {E}^{(n)}$ if $h$ is subadditive.

\begin{theorem}\label{mem'}
	Let ${E'}^{(n)}$ be a nonnegative function defined as in Eq.~\eqref{max1}. If it is nonincreasing on average under LOCC, then (i) ${E'}^{(3)}$ is a complete MEM but not tightly complete monogamous, and if $h$ is strictly concave, ${E'}^{(3)}$ is completely monogamous, and
	(ii) ${E'}^{(n)}$ is not complete whenever $n\geq4$.
\end{theorem}

\begin{proof}
	(i) It is clear that the unification condition and the hierarchy condition are valid for ${E'}^{(3)}$, thus ${E'}^{(3)}$ is a complete MEM. Let ${E'}^{(3)}(|\psi\ra^{ABC})={E'}^{(2)}(|\psi\ra^{A|BC})$, then $\rho^{BC}$ is not necessarily separable. Thus, ${E'}^{(3)}$ is not tightly complete monogamous. If $h$ is strictly concave and ${E'}^{(3)}(|\psi\ra^{ABC})={E'}^{(2)}(\rho^{AB})$,
	then ${E'}^{(2)}(|\psi\ra^{A|BC})={E'}^{(2)}(|\psi\ra^{B|AC})={E'}^{(2)}(\rho^{AB})$. Therefore, $|\psi\ra^{ABC}=|\psi\ra^{AB}|\psi\ra^C$ by Theorem in Ref.~\cite{GG2019}. That is, ${E'}^{(3)}$ is completely monogamous.
		
	(ii) Let 
	\bea
	|W_4\ra=\frac12\left(|1000\ra+|0100\ra+|0010\ra+|0001\ra\right), 
	\eea
	we have ${E'}^{(4)}(|W_4\ra)<{E'}^{(2)}(|W_4\ra^{AB|CD})$ since
	\beax
	\rho^A=\rho^B=\rho^C=\rho^D=\left(\begin{array}{cc}
		3/4&0\\
		0&1/4	
	\end{array} \right) 
	\eeax
	and the bipartite reduced state is maximal mixed two qubit state. That is, it violates the hierarchy condition. This complete the proof.
\end{proof}

We denote the corresponding ${E'}^{(n)}$ in the previous subsection by ${E'}_f^{(n)}$, ${C'}^{(n)}$, ${\tau'}^{(n)}$, ${E'}_q^{(n)}$, ${E'}_\alpha^{(n)}$, ${N'}_F^{(n)}$, ${E'}_{\mathcal{F}}^{(n)}$, ${E'}_{\mathcal{F}'}^{(n)}$, ${E'}_{A\mathcal{F}}^{(n)}$,  ${E'}_2^{(n)}$, ${E'}_{\min}^{(n)}$, ${E'}_{\min'}^{(n)}$, and ${\hat{N'}}^{(n)}$, respectively. Note here that we can not prove here ${E'}^{(n)}$ is nonincreasing on average under LOCC, but we conjecture that ${E'}^{(n)}$ does not increase on average under LOCC for these cases mentioned above. Hereafter we always assume the conjecture is true. 
In such a sense, by Theorem~\ref{mem'}, all of them are complete MEMs but not tightly complete monogamous by Theorem~\ref{mem'} for the case of $n=3$, and ${E'}_f^{(3)}$, ${C'}^{(3)}$, ${\tau'}^{(3)}$, ${E'}_q^{(3)}$, ${E'}_\alpha^{(3)}$, ${N'}_F^{(3)}$, ${E'}_{\mathcal{F}}^{(3)}$, ${E'}_{\mathcal{F}'}^{(3)}$, and ${E'}_{A\mathcal{F}}^{(3)}$ are completely monogamous, ${E'}_f^{(n)}$, ${C'}^{(n)}$, ${\tau'}^{(n)}$, ${E'}_q^{(n)}$, ${E'}_\alpha^{(n)}$, ${N'}_F^{(n)}$, ${E'}_{\mathcal{F}}^{(n)}$, ${E'}_{\mathcal{F}'}^{(n)}$, ${E'}_{A\mathcal{F}}^{(n)}$, ${E'}_2^{(n)}$, ${E'}_{\min}^{(n)}$, ${E'}_{\min'}^{(n)}$, and ${\hat{N'}}^{(n)}$ are not complete MEMs whenever $n\geq4$.

If $h$ is not strictly concave, then ${E'}^{(3)}$ is not completely monogamous. For example, we take
\bea\label{eg1}
|\psi\ra^{ABC}=|\psi\ra^{AB_1}|\psi\ra^{B_2C},
\eea
where $B_1B_2$ means $\mH^B$ has a subspace isomorphic to $\mH^{B_1}\otimes\mH^{B_2}$ and up to local unitary on system $B_1B_2$. We assume 
\beax {E'}_{\min}^{(3)}(|\psi\ra^{ABC})={E'}_{\min}^{(2)}(|\psi\ra^{A|BC}),\\ {\hat{N'}}^{(3)}(|\psi\ra^{ABC})={\hat{N'}}^{(2)}(|\psi\ra^{A|BC}),~
\eeax
then 
\beax {E'}_{\min}^{(3)}(|\psi\ra^{ABC})={E'}_{\min}^{(2)}(|\psi\ra^{AB_1})={E'}_{\min}^{(2)}(\rho^{AB}),\\
{\hat{N'}}^{(3)}(|\psi\ra^{ABC})={\hat{N'}}^{(2)}(|\psi\ra^{AB_1})={\hat{N'}}^{(2)}(\rho^{AB}),~
\eeax 
and $\rho^{BC}$ is entangled. In addition, we take
\bea\label{eg2}
|\phi\rangle^{ABC} =\frac{1}{\sqrt{3}}|000\rangle+\frac{1}{\sqrt{3}}|101\rangle+\frac{1}{\sqrt{3}}| 110\rangle.
\eea
It is straightforward that
\beax
&&{E'}_{2}^{(3)}(|\phi\rangle^{ABC})\\
&=&{E'}_{2}^{(2)}(|\phi\rangle^{A|BC})={E'}_{2}^{(2)}(|\phi\rangle^{AB|C})
={E'}_{2}^{(2)}(|\phi\rangle^{B|AC})\\
&=&{E'}_{2}^{(2)}(\rho^{AB})={E'}_{2}^{(2)}(\rho^{AC})
={E'}_{2}^{(2)}(\rho^{BC})\\
&=&1/3.
\eeax
Namely, ${E'}_2^{(3)}$, ${E'}_{\min}^{(3)}$, ${E'}_{\min'}^{(3)}$, and ${\hat{N'}}^{(3)}$ are not completely monogamous. Namely, for these four complete MEMs, monogamy coincides with complete monogamy, tightly complete monogamy seems stronger than both monogamy and complete monogamy.

It is worthy mentioning here that ${E'}^{(n)}$ may not a unified MEM if $n\geq 4$ since it may occur that
${E'}^{(k)}(X_1|X_2| \cdots| X_{k})<{E'}^{(l)}(Y_1|Y_2| \cdots |Y_{l})$ for some state $\rho\in\mathcal{S}^{A_1A_2\cdots A_n}$ whenever $X_1|X_2| \cdots| X_{k}\succ^a Y_1|Y_2| \cdots |Y_{l}$.

Let $|\psi\ra^{A_1A_2\cdots A_n}$ be a pure state in $\mH^{A_1A_2\cdots A_n}$ and $h$ be a non-negative concave function on $\mS^X$. We define
\bea\label{max2}
{\mE'}^{(n)}(|\psi\ra^{A_1A_2\cdots A_n})
=\max_{i_1\leq \cdots\leq i_s, s\leq n/2}h(\rho^{A_{i_1}A_{i_2}\cdots A_{i_s}})\quad\quad
\eea
for pure states and for mixed states by the convex-roof structure. By definition, 
\bea
{E'}^{(n)}\leq{\mE'}^{(n)},
\eea
${\mE'}^{(3)}$ coincides with ${E'}^{(3)}$, ${\mE'}^{(n)}$ satisfies the hierarchy condition, but it may violate
the unification condition. We give a comparison for these MEMs in Table~\ref{tab:table3} for more clarity.

\begin{table}
	\caption{\label{tab:table3} Comparing of ${E'}^{(3)}$ with different reduced functions, ${E'}^{(4)}$~($n\geq4$), and ${\mE'}^{(4)}$~($n\geq4$).}	
	\begin{ruledtabular}
		\begin{tabular}{ccccc}
			MEM     & Unified & Complete & CM & TCM \\ \colrule
			${E'}_f^{(3)}$          &$\checkmark$&$\checkmark$ &$\checkmark$&$\times$\\
			${C'}^{(3)}$            &$\checkmark$&$\checkmark$ &$\checkmark$&$\times$\\
			${\tau'}^{(3)}$         &$\checkmark$&$\checkmark$ &$\checkmark$&$\times$   \\
			${E'}_q^{(3)}$          &$\checkmark$&$\checkmark$ &$\checkmark$&$\times$\\
			${E'}_\alpha^{(3)}$     &$\checkmark$&$\checkmark$ &$\checkmark$&$\times$\\
			${N'}_F^{(3)}$          &$\checkmark$&$\checkmark$ &$\checkmark$&$\times$\\
			${E'}_{\mF}^{(3)}$      &$\checkmark$&$\checkmark$ &$\checkmark$&$\times$\\
			${E'}_{\mF'}^{(3)}$     &$\checkmark$&$\checkmark$ &$\checkmark$&$\times$\\
			${E'}_{A\mF}^{(3)}$     &$\checkmark$&$\checkmark$ &$\checkmark$&$\times$\\
			${E'}_{2}^{(3)}$        &$\checkmark$&$\checkmark$ &$\times$    &$\times$\\
			${E'}_{\min}^{(3)}$     &$\checkmark$&$\checkmark$ &$\times$    &$\times$\\
			${E'}_{\min'}^{(3)}$    &$\checkmark$&$\checkmark$ &$\times$    &$\times$\\
			${\hat{N'}}^{(3)}$      &$\checkmark$&$\checkmark$ &$\times$    &$\times$\\  		
			${E'}^{(4)}$~($n\geq4$)         &?&$\times$ &?&$\times$\\	
			${\mE'}^{(n)}$~($n\geq4$)       &?&$\times$ &?&$\times$
		\end{tabular}
	\end{ruledtabular}
\end{table}

The case ${E'}^{(n)}<{\mE'}^{(n)}$ occurs whenever $n\geq4$. It is clear that ${E'}^{(4)}(|W_4\ra)<{\mE'}^{(4)}(|W_4\ra)$ for any ${E'}^{(4)}$ and ${\mE'}^{(4)}$ with the reduced functions we considered in Sec. III. In addition, for the state
\bea\label{eg1-2}
|\varphi\ra=\frac{\sqrt{5}}{4}|0000\ra+\frac{\sqrt{5}}{4}|1111\ra+\frac14|0100\ra+\frac{\sqrt{5}}{4}|1010\ra,~~~~~~
\eea
we have ${E'}^{(4)}<{\mE'}^{(4)}$ for any ${E'}^{(4)}$ and ${\mE'}^{(4)}$ mentioned above except for $h=h_{\min}$ and
$h=\hat{h}$ since $\rho^A=\rho^B=\rho^C=\rho^D$ and the eigenvalues of $\rho^A$ is $\{5/8, 3/8\}$ and the eigenvalues of the bipartite reduced state is $\{3/8, 5/16, 5/16\}$.

\section{Complete GMEM}

\subsection{Complete GMEM from sum of the reduced functions}

In Ref.~\cite{Guo2022entropy}, we discussed the completeness of GMEMs defined by sum of all reduced functions of the single subsystems with the reduced functions corresponding to $E_f$, $C$, $\tau$, $E_q$, and $E_\alpha$. We consider here the general case for any given bipartite entanglement monotone. Let $|\psi\ra^{A_1A_2\cdots A_n}$ be a pure state in $\mH^{A_1A_2\cdots A_n}$ and $h$ be a non-negative concave function on $\mS^X$. We define
\begin{widetext}
\bea\label{gsum1}
E_g^{(n)}(|\psi\ra^{A_1A_2\cdots A_n})
=\begin{cases}
	\dfrac{1}{2}\sum_ih(\rho^{A_i}), & h(\rho^X)>0 ~\text{for any bipartition $X|Y$ of $A_1A_2\cdots A_n$},\\
	0, & \text{otherwise}
\end{cases}\quad
\eea
\end{widetext}
and then extend it to mixed states by the convex-roof structure. By Proposition 1 and Proposition 4 in Ref.~\cite{Guo2022entropy}, together with Theorem~\ref{th1}, we have the following statement.

\begin{pro}\label{gmem1}
	Let $E_g^{(n)}$ be a non-negative function defined as in Eq.~\eqref{gsum1}. Then the following statements hold true.
	\begin{itemize}
		\item[(i)] $E_g^{(n)}$ is a unified GMEM and is completely monogamous;
		\item[(ii)] $E_g^{(n)}$ is a complete GMEM iff $h$ is subadditive;
		\item[(iii)] $E_g^{(n)}$ is tightly complete monogamous iff $h$ is subadditive with Eq.~\eqref{sa-h} holds.
	\end{itemize}
\end{pro}

We denote $E_g^{(n)}$ in the previous Section by $E_{g,f}^{(n)}$, $C_g^{(n)}$, $\tau_g^{(n)}$, $E_{g,q}^{(n)}$, $E_{g,\alpha}^{(n)}$, $N_{g,F}^{(n)}$, $E_{g,\mathcal{F}}^{(n)}$, $E_{g,\mathcal{F}'}^{(n)}$, $E_{g,A\mathcal{F}}^{(n)}$, $E_{g,2}^{(n)}$, $E_{g,\min}^{(n)}$, ${E}_{g,\min'}^{(n)}$, and $\hat{N}_g^{(n)}$, respectively. By Proposition~\ref{gmem1},
we can conclude:
(i) All these GMEMs are unified GMEMs and are completely monogamous;
(ii) $E_{g,f}^{(n)}$, $C_g^{(n)}$, $\tau_g^{(n)}$, $E_{g,q}^{(n)}$, 
$E_{g,\mathcal{F}}^{(n)}$, $E_{g,\mathcal{F}'}^{(n)}$, $E_{g,A\mathcal{F}}^{(n)}$, 
$E_{g,2}^{(n)}$, and $\hat{N}_g^{(n)}$ are complete GMEMs;
(iii) $E_{g,\alpha}^{(n)}$, $N_{g,F}^{(n)}$, $E_{g,\min}^{(n)}$, and ${E}_{g,\min'}^{(n)}$ are not complete GMEMs since the associated reduced functions are not subadditive and thus they violate the hierarchy condition for some states.
(iv) $E_{g,f}^{(n)}$, $C_g^{(n)}$, $\tau_g^{(n)}$, and 
$E_{g,2}^{(n)}$ are tightly complete monogamous. 
Therefore, for these GMEMs, tightly complete monogamy are stronger than the complete monogamy under the frame work of the complete GMEM.

By the assumption, we conjecture that $E_{g,q}^{(n)}$, $E_{g,\mathcal{F}}^{(n)}$, $E_{g,\mathcal{F}'}^{(n)}$, $E_{g,A\mathcal{F}}^{(n)}$, and $\hat{N}_g^{(n)}$ are tightly complete monogamous. That is, $E_g^{(n)}$ is complete, completely monogamous, tightly complete monogamous, if and only if $E^{(n)}$ is complete, completely monogamous, tightly complete monogamous, respectively.

A similar quantity, $\varepsilon_{g-12\cdots n(2)}$, is also put forward in Ref.~\cite{Guo2022jpa}. 
Let $|\psi\ra^{A_1A_2\cdots A_n}$ be a pure state in $\mH^{A_1A_2\cdots A_n}$ and $h$ be a non-negative concave function on $\mS^X$. We define
\begin{widetext}
\bea\label{gsum2}
\mE_g^{(n)}(|\psi\ra^{A_1A_2\cdots A_n})
=\begin{cases}
	\mE^{(n)}(|\psi\ra^{A_1A_2\cdots A_n}), & h(\rho^X)>0 ~\text{for any bipartition $X|Y$ of $A_1A_2\cdots A_n$},\\
	0, & \text{otherwise},
\end{cases}
\eea
\end{widetext}
and then extend it to mixed states by the convex-roof structure. Notice here that $\mE_g^{(n)}$ is slightly different than $\varepsilon_{g-12\cdots n(2)}$ in which the factor ``$1/2$'' is ignored.

Clearly, 
\bea
{E}_g^{(n)}\leq{\mE}_g^{(n)},
\eea
and ${\mE}_g^{(3)}$ coincides with ${E}_g^{(3)}$ but ${\mE}_g^{(n)}$ is different from ${E}_g^{(n)}$ whenever $n\geq4$.
${\mE}_g^{(n)}$ is just ${\mE}_{g-12\cdots n(2)}$ in Ref.~\cite{Guo2022jpa} if the corresponding bipartite entanglement measure is an entanglement monotone. The following Proposition can be easily checked.

\begin{table}
	\caption{\label{tab:table4} Comparing of $E_{g}^{(n)}$ with different reduced functions and $\mE_{g}^{(n)}$ ($n\geq4$).}	
	\begin{ruledtabular}
		\begin{tabular}{ccccc}
			GMEM                 & Unified    & Complete    & CM         & TCM   \\ \colrule
			$E_{g,f}^{(n)}$      &$\checkmark$&$\checkmark$ &$\checkmark$&$\checkmark$\\
			$C_g^{(n)}$          &$\checkmark$&$\checkmark$ &$\checkmark$&$\checkmark$\\
			$\tau_g^{(n)}$       &$\checkmark$&$\checkmark$ &$\checkmark$&$\checkmark$\\
			$E_{g,q}^{(n)}$      &$\checkmark$&$\checkmark$ &$\checkmark$&$\checkmark\footnotemark[1]$ \\
			$E_{g,\alpha}^{(n)}$ &$\checkmark$&$\times$     &$\checkmark$&$\times$     \\
			$N_{g,F}^{(n)}$      &$\checkmark$&$\times$     &$\checkmark$&$\times$     \\
			$E_{g,\mF}^{(n)}$    &$\checkmark$&$\checkmark$ &$\checkmark$&$\checkmark\footnotemark[1]$\\
			$E_{g,\mF'}^{(n)}$   &$\checkmark$&$\checkmark\footnotemark[2]$ &$\checkmark$&$\checkmark\footnotemark[1]$\\
			$E_{g,A\mF}^{(n)}$   &$\checkmark$&$\checkmark\footnotemark[2]$&$\checkmark$&$\checkmark\footnotemark[1]$\\
			$E_{g,2}^{(n)}$      &$\checkmark$&$\checkmark$ &$\checkmark$&$\checkmark$\\
			$E_{g,\min}^{(n)}$   &$\checkmark$&$\times$     &$\checkmark$&$\times$    \\
			${E}_{g,\min'}^{(n)}$&$\checkmark$&$\times$     &$\checkmark$&$\times$    \\
			${\hat{N}}_g^{(n)}$  &$\checkmark$&$\checkmark\footnotemark[2]$&$\checkmark$&$\checkmark\footnotemark[1]$\\ 
			$\mE_{g}^{(n)}$~$(n\geq4)$     &$\checkmark$&$\checkmark$ &$\checkmark$&$\checkmark$
		\end{tabular}
	\end{ruledtabular}
	\footnotetext[1]{Assume that $h$ is subadditive and Eq.~\eqref{sa-h} holds.}\\
	\footnotetext[2]{Assume that $h$ is subadditive.}
\end{table}

\begin{pro}\label{th22}
	Let $\mE^{(n)}$ be a non-negative function defined as in Eq.~\eqref{gsum2}, $n\geq4$. Then $\mE^{(n)}$ is a complete MEM and it is completely monogamous and tightly complete monogamous.
\end{pro}

That is, for the case of $n\geq4$, all these MEMs $\mE_{g}^{(n)}$ with the reduced functions we discussed in Sec. III are complete GMEMs, and are not only completely monogamous but also tightly complete monogamous. For convenience, we list all these MEMs in Table~\ref{tab:table4}. In addition, it is obvious that $E_g^{(n)}<\mE_g^{(n)}$ whenever $n\geq4$ for any $E_g^{(4)}$ and $\mE_g^{(4)}$ mentioned above.

\subsection{Complete GMEM from the maximal reduced function}

Let $|\psi\ra^{A_1A_2\cdots A_n}$ be a pure state in $\mH^{A_1A_2\cdots A_n}$ and $h$ be a non-negative concave function on the set of density matrices. We define
\begin{widetext}
\bea\label{gmax1}
E_{g'}^{(n)}(|\psi\ra^{A_1A_2\cdots A_n})
=\begin{cases}
	\max\limits_ih(\rho^{A_i}),& h(\rho^X)>0 ~\text{for any bipartition $X|Y$ of $A_1A_2\cdots A_n$},\\
	0, & \text{otherwise},
	\end{cases} \quad 
\eea
\end{widetext}
and then extend it to mixed states by the convex-roof structure. From Theorem~\ref{mem'}, we have the following Proposition.

\begin{pro}\label{gmem'}
	Let ${E}_{g'}^{(n)}$ be a nonnegative function defined as in Eq.~\eqref{gmax1}. If it is nonincreasing on average under LOCC, then (i) ${E}_{g'}^{(3)}$ is a complete GMEM but not tightly complete monogamous, and if $h$ is strictly concave, ${E}_{g'}^{(3)}$ is completely monogamous, and
	(ii) ${E}_{g'}^{(n)}$ is not complete whenever $n\geq4$.
\end{pro}

We denote $E_{g'}^{(n)}$ the corresponding GMEMs mentioned in the previous Subsection by $E_{g',f}^{(n)}$, $C_{g'}^{(n)}$, $\tau_{g'}^{(n)}$, $E_{g',q}^{(n)}$, $E_{g',\alpha}^{(n)}$, $N_{g',F}^{(n)}$,
$E_{g',\mathcal{F}}^{(n)}$, $E_{g',\mathcal{F}'}^{(n)}$, $E_{g',A\mathcal{F}}^{(n)}$, $E_{g',2}^{(n)}$, $E_{g',\min}^{(n)}$, ${E}_{g',\min'}^{(n)}$, and $\hat{N}_{g'}^{(n)}$, respectively. Then all these GMEMS are complete GMEMs but not tightly complete monogamous for the case of $n=3$, $E_{g',f}^{(3)}$, $C_{g'}^{(3)}$, $\tau_{g'}^{(3)}$, $E_{g',q}^{(3)}$, $E_{g',\alpha}^{(3)}$, $N_{g',F}^{(3)}$, $E_{g',\mathcal{F}}^{(3)}$, $E_{g',\mathcal{F}'}^{(3)}$, and $E_{g',A\mathcal{F}}^{(3)}$, are completely monogamous, all of these GMEMs are not complete GMEMs whenever $n\geq4$.

One need note here that, when $h$ is not strictly concave, $E_{g'}^{(n)}$ is not a unified GMEM since it may happen that ${E}_{g'}^{(k)}(X_1|X_2|\cdots|X_{k})= {E}_{g'}^{(l)}(Y_1|Y_2|\cdots|Y_{l})$ for some $\rho^{A_1A_2\cdots A_n}\in\mS_g^{A_1A_2\cdots A_n}$ with $X_1|X_2| \cdots| X_{k}\succ^a Y_1|Y_2| \cdots |Y_{l}$, namely, it violates Eq.~\eqref{gcoarsen}. In addition, $E_{g'}^{(n)}$ also violates Eq.~\eqref{g-tight1} or Eq.~\eqref{g-tight2}.
For example, we take the state in Eq.~\eqref{eg1} with both $|\psi\ra^{AB_1}$ and $|\psi\ra^{B_2C}$ are entangled. 
We assume 
\beax {E}_{g',\min}^{(3)}(|\psi\ra^{ABC})={E}_{g',\min}^{(2)}(|\psi\ra^{A|BC}),\\ {\hat{N}}_{g'}^{(3)}(|\psi\ra^{ABC})={\hat{N}}_{g'}^{(2)}(|\psi\ra^{A|BC}),~~
\eeax
then 
\beax {E}_{g',\min}^{(3)}(|\psi\ra^{ABC})={E}_{g',\min}^{(2)}(|\psi\ra^{AB_1})={E}_{g',\min}^{(2)}(\rho^{AB}),\\
{\hat{N}}_{g'}^{(3)}(|\psi\ra^{ABC})={\hat{N}}_{g'}^{(2)}(|\psi\ra^{AB_1})={\hat{N}}_{g'}^{(2)}(\rho^{AB}),~~~
\eeax 
and $\rho^{BC}$ is entangled. In addition, for the state in Eq.~\eqref{eg2}, we have
\beax
&&E_{g',2}^{(3)}(|\phi\rangle^{ABC})\\
&=&E_{g',2}^{(2)}(|\phi\rangle^{A|BC})=E_{g',2}^{(2)}(|\phi\rangle^{AB|C})=E_{g',2}^{(2)}(|\phi\rangle^{B|AC})\\
&=&E_{g',2}^{(2)}(\rho^{AB})=E_{g',2}^{(2)}(\rho^{AC})=E_{g',2}^{(2)}(\rho^{BC})\\
&=&\frac13.
\eeax
That is, whenever $h$ is strictly concave, $E_{g'}^{(n)}$ is complete, completely monogamous, tightly complete monogamous, if and only if ${E'}^{(n)}$ is complete, completely monogamous, tightly complete monogamous, respectively.

\begin{table}
	\caption{\label{tab:table5} Comparing of $E_{g'}^{(3)}$ with different reduced functions, $E_{g'}^{(4)}$, and $\mE_{g'}^{(n)}$.}	
	\begin{ruledtabular}
		\begin{tabular}{ccccc}
			GMEM                 & Unified    & Complete    & CM         & TCM   \\ \colrule			
			$E_{g',f}^{(3)}$     &$\checkmark$&$\checkmark$ &$\checkmark$&$\times$\\
			$C_{g'}^{(3)}$       &$\checkmark$&$\checkmark$ &$\checkmark$&$\times$\\
			$\tau_{g'}^{(3)}$    &$\checkmark$&$\checkmark$ &$\checkmark$ &$\times$\\
			$E_{g',q}^{(3)}$     &$\checkmark$&$\checkmark$ &$\checkmark$ &$\times$\\
			$E_{g',\alpha}^{(3)}$&$\checkmark$&$\checkmark$ &$\checkmark$ &$\times$\\
			$N_{g',F}^{(3)}$     &$\checkmark$&$\checkmark$ &$\checkmark$ &$\times$\\
			$E_{g',\mF}^{(3)}$   &$\checkmark$&$\checkmark$ &$\checkmark$ &$\times$\\
			$E_{g',\mF'}^{(3)}$  &$\checkmark$&$\checkmark$ &$\checkmark$ &$\times$\\
			$E_{g',A\mF}^{(3)}$  &$\checkmark$&$\checkmark$ &$\checkmark$ &$\times$\\
			$E_{g',2}^{(3)}$     &$\times$&$\times$&$\times$&$\times$\\
			$E_{g',\min}^{(3)}$  &$\times$&$\times$&$\times$&$\times$\\
			${E}_{g',\min'}^{(3)}$&$\times$&$\times$&$\times$&$\times$\\
			${\hat{N}}_{g'}^{(3)}$&$\times$&$\times$&$\times$&$\times$\\\colrule			
			$E_{g'}^{(n)}$~$(n\geq4)$     &?&$\times$ &?&$\times$\\						
			$\mE_{g'}^{(n)}$~$(n\geq4)$    &?&$\times$ &?&$\times$
		\end{tabular}
	\end{ruledtabular}
\end{table}

For the states that admit the form 
\bea\label{eta}
|\eta\ra^{ABC}=|\eta\ra^{AB_1}|\eta\ra^{B_2C}
\eea
where $B_1B_2$ refers to $\mH^B$ has a subspace isomorphic to $\mH^{B_1^{(x)}}\otimes\mH^{B_2^{(x)}}$ such that up to local unitary on system $B$, we have $E_{g'}^{(3)}(|\eta\ra^{ABC})=E_{g'}^{(2)}(|\eta\ra^{B|AC})$ whenever $h(\rho\ot\sigma)\geq h(\rho)$ and $h(\rho\ot\sigma)\geq h(\sigma)$ for any $\rho$ and $\sigma$, and $\rho^{AC}$ is a product state.  We therefore have the following fact.

\begin{pro}\label{gmem'2}
	If $h$ is strictly concave and $h(\rho\ot\sigma)\geq h(\rho)$ and $h(\rho\ot\sigma)\geq h(\sigma)$ for any $\rho$ and $\sigma$, $E_{g'}^{(3)}$ defined as in Eq.~\eqref{gmax1} is tightly complete monogamous on the states that admit the form~\eqref{eta}. 
\end{pro}

In fact, we always have $h(\rho\ot\sigma)\geq h(\rho)$ and $h(\rho\ot\sigma)\geq h(\rho)$ if $h\in\{S$, $h_C$, $h_\tau$, $h_q$, $h_\alpha$, $h_N$, $h_{\mathcal{F}}$, $h_{\mathcal{F}'}$, $h_{A\mathcal{F}}\}$.
So $E_{g',f}^{(n)}$, $C_{g'}^{(n)}$, $\tau_{g'}^{(n)}$, $E_{g',q}^{(n)}$, $E_{g',\alpha}^{(n)}$, $N_{g',F}^{(n)}$,
$E_{g',\mathcal{F}}^{(n)}$, $E_{g',\mathcal{F}'}^{(n)}$, and $E_{g',A\mathcal{F}}^{(n)}$ are tightly complete monogamous on the states with the form as in Eq.~\eqref{eta}. Proposition~\ref{gmem'2} is also valid when we replacing $E_{g'}^{(3)}$ with ${E'}^{(3)}$.

Let $|\psi\ra^{A_1A_2\cdots A_n}$ be a pure state in $\mH^{A_1A_2\cdots A_n}$ and $h$ be a non-negative concave function on $\mS^X$. We define
\begin{widetext}
\bea\label{gmax2}
{\mE}_{g'}^{(n)}(|\psi\ra^{A_1A_2\cdots A_n})
=\begin{cases}
	{\mE'}^{(n)}(|\psi\ra^{A_1A_2\cdots A_n}),& h(\rho^X)>0 ~\text{for any bipartition $X|Y$ of $A_1A_2\cdots A_n$},\\
	0, & \text{otherwise},
\end{cases}
\eea
\end{widetext}
for pure states and for mixed states by the convex-roof structure. By definition, 
\bea
{E}_{g'}^{(n)}\leq{\mE}_{g'}^{(n)},
\eea
${\mE}_{g'}^{(3)}$ coincides with ${E'}_g^{(3)}$, and ${\mE}_{g'}^{(n)}$ satisfies the hierarchy condition, but it violates the unification condition if $n\geq4$. It is easy to see that all these GMEMs ${{\mE}}_{g'}^{(n)}$ with the reduced function we discussed are not complete GMEMs whenever $n\geq4$. 
We give comparison for these GMEMs in Table~\ref{tab:table5}.

For the case of $n\geq4$, it is possible that ${E}_{g'}^{(n)}<{\mE}_{g'}^{(n)}$. For example, for $|W_4\ra$ and the state in Eq.~\eqref{eg1-2} we have ${E}_{g'}^{(4)}<{\mE}_{g'}^{(4)}$ for any ${E}_{g'}^{(4)}$ and ${\mE}_{g'}^{(4)}$ mentioned above except for $h=h_{\min}$ and $h=\hat{h}$.

\subsection{GMEM from the minimal reduced function}

With $h$ is a non-negative concave function on the set of density matrices, when we define
\begin{widetext}
\bea\label{gmin1}
E_{g''}^{(n)}(|\psi\ra^{A_1A_2\cdots A_n})
=\begin{cases} \min\limits_ih(\rho^{A_i}),& h(\rho^X)>0 ~\text{for any bipartition $X|Y$ of $A_1A_2\cdots A_n$},\\
	0, & \text{otherwise},
\end{cases} 
\eea
\end{widetext}
and then extend it to mixed states by the convex-roof structure, it is a GMEM. Moreover, we can define
\bea\label{gmin2}
\mE_{g''}^{(n)}(|\psi\ra^{A_1A_2\cdots A_n})
=\min_{i_1\leq \cdots\leq i_s, s\leq n/2}h(\rho^{A_{i_1}A_{i_2}\cdots A_{i_s}}),\quad~
\eea
and then extend it to mixed states by the convex-roof structure, it is also a GMEM. For example, GMC, denoted by $C_{gme}$~\cite{Ma2011}, is defined as in Eq.~\eqref{gmin2}. Recall that,
\beax
C_{gme}(|\psi\ra):=\min\limits_{\gamma_i \in \gamma} \sqrt{2\left[ 1-\tr(\rho^{A_{\gamma_i}})^{2}\right] }
\eeax
for pure state $|\psi\ra\in\mH^{A_1A_2\cdots A_m}$, where $\gamma=\{\gamma_i\}$ represents the set of all possible bipartitions of $A_1A_2\cdots A_m$, and via the convex-roof extension for mixed states.

\begin{figure}	
	\centering
	\subfigure[$C_g$]{
		\hspace{-28mm}\begin{minipage}[c]{0.3\textwidth}
			\centering
			\includegraphics[width=78mm]{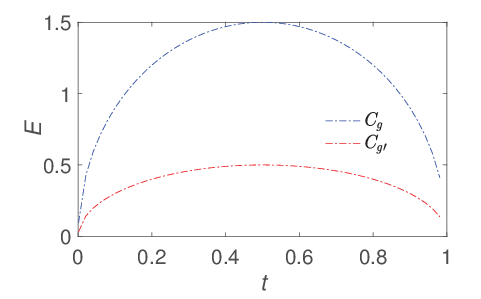}
		\end{minipage}
	}	\\
	\subfigure[$E_{g,\mathcal{F}'}$]{
		\hspace{-28mm}\begin{minipage}[c]{0.3\textwidth}
			\centering
			\includegraphics[width=78mm]{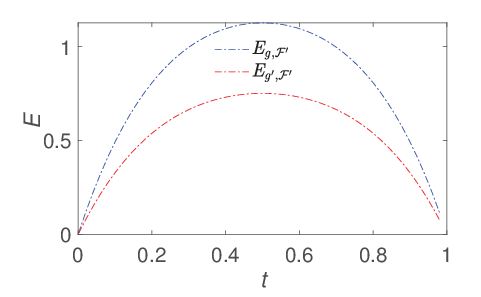}
		\end{minipage}
	}	 \\
	
	\subfigure[$E_{g,2}$]{
		\hspace{-28mm}\begin{minipage}[c]{0.3\textwidth}
			\centering
			\includegraphics[width=78mm]{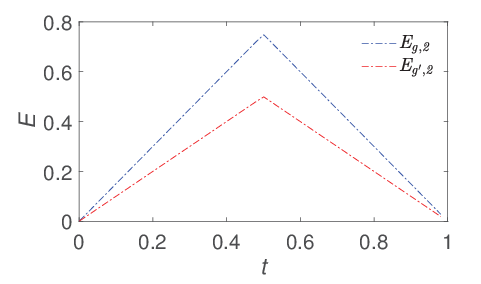}
		\end{minipage}
	}	
	\caption{\label{fig1}(color online). Comparing (a) $C_g^{(3)}$ and $C_{g'}^{(3)}$,
		(b) $E_{g,\mathcal{F}'}^{(3)}$ and $E_{{g',\mathcal{F}'}}^{(3)}$, and (c) $E_{g,2}^{(3)}$ and $E_{g',2}^{(3)}$ for $|\Psi\ra$, respectively. $E_{g'}^{(3)}=E_{g''}^{(3)}$ in such a case.}
\end{figure}

We denote $E_{g''}^{(n)}$ the corresponding GMEMs mentioned in the previous Subsection by $E_{g'',f}^{(n)}$, $C_{g''}^{(n)}$, $\tau_{g''}^{(n)}$, $E_{g'',q}^{(n)}$, $E_{g'',\alpha}^{(n)}$, $N_{g'',F}^{(n)}$,
$E_{g'',\mathcal{F}}^{(n)}$, $E_{g'',\mathcal{F}'}^{(n)}$, $E_{g'',A\mathcal{F}}^{(n)}$, 
$E_{g'',2}^{(n)}$, $E_{g'',\min}^{(n)}$, ${E}_{g'',\min'}^{(n)}$, and $\hat{N}_{g''}^{(n)}$, respectively,
and denote ${{\mE}}_{g''}^{(n)}$ by ${{\mE}}_{g'',f}^{(n)}$, ${{\mC}}_{g''}^{(n)}$ (or $C_{gme}$), ${{\hat{\tau}}}_{g''}^{(n)}$, ${{\mE}}_{g'',q}^{(n)}$, ${{\mE}}_{g'',\alpha}^{(n)}$, ${{\mN}}_{g'',F}^{(n)}$,
${{\mE}}_{g'',\mathcal{F}}^{(n)}$, ${{\mE}}_{g'',\mathcal{F}'}^{(n)}$, ${{\mE}}_{g'',A\mathcal{F}}^{(n)}$, 
${{\mE}}_{g'',2}^{(n)}$, ${{\mE}}_{g'',\min}^{(n)}$, ${{\mE}}_{g'',\min'}^{(n)}$, and ${\hat{\mN}}_{g''}^{(n)}$, respectively.

By definition,
\bea
\mE_{g''}^{(n)}\leq E_{g''}^{(n)}\leq E_{g'}^{(n)}\leq E_{g}^{(n)}
\eea
for any $h$, and $\mE_{g''}^{(3)}=E_{g''}^{(3)}$. If $n\geq4$, there does exist state such that $\mE_{g''}^{(n)}< E_{g''}^{(n)}$. For example, we take
\beax
|\psi\ra^{ABCD}=|\psi\ra^{AB_1}|\psi\ra^{B_2C_1}|\psi\ra^{C_2D},
\eeax
where $X_1X_2$ refers to $\mH^X$ has a subspace isomorphic to $\mH^{X_1^{(x)}}\otimes\mH^{X_2^{(x)}}$ such that up to local unitary on system $X$. If $h(\rho^{B_2})<h(\rho^A)$ and $h(\rho^{B_2})<h(\rho^D)$, then $\mE_{g''}^{(4)}(|\psi\ra^{ABCD})=h(\rho^{B_2})< E_{g''}^{(4)}(|\psi\ra^{ABCD})$. In addition, for the state in Eq.~\eqref{eg1-2},
\beax
\mE_{g'',\min}^{(4)}=\frac{5}{16}<{E}_{g'',\min}^{(4)}=\frac38,\\
\hat{\mN}_{g''}^{(4)}=\dfrac{\sqrt{15}}{8\sqrt{2}}<\hat{N}_{g''}^{(4)}=\frac{\sqrt{15}}{8}.
\eeax

$C_{gme}$ is not a complete GMEM since it does not satisfy the hierarchy condition~\eqref{ghierarchy}~\cite{Guo2022entropy}: Let 
\bea\label{eg3}
|\xi\ra=\frac{\sqrt{5}}{4}|0000\ra+\frac14|1111\ra+\frac{\sqrt{5}}{4}|0100\ra+\frac{\sqrt{5}}{4}|1010\ra,~~~~~~
\eea
then
\beax
C_{gme}(|\xi\ra)&=&C(|\xi\ra^{ABC|D})=\frac{\sqrt{15}}{8}\\
&<&C(|\xi\ra^{AB|CD})=\frac{\sqrt{65}}{8}.
\eeax

In general, $E_{g''}^{(n)}$ and $\mE_{g''}^{(n)}$ do not obey the unification condition~\eqref{gcoarsen} and the hierarchy condition~\eqref{ghierarchy}. For instance, for the state as in Eq.~\eqref{eg1}, we have 
\beax {E}_{g'',\min}^{(3)}(|\psi\ra^{ABC})={E}_{g'',\min}^{(2)}(|\psi\ra^{B|AC}),\\ {\hat{N}}_{g''}^{(3)}(|\psi\ra^{ABC})={\hat{N}}_{g''}^{(2)}(|\psi\ra^{B|AC}),
\eeax
and 
\beax &&{E}_{g'',\min}^{(3)}(|\psi\ra^{ABC})<{E}_{g'',\min}^{(2)}(|\psi\ra^{A|BC})\\
&=&{E}_{g'',\min}^{(2)}(|\psi\ra^{AB_1})={E}_{g'',\min}^{(2)}(\rho^{AB}),
\eeax
\beax&&{E}_{g'',\min}^{(3)}(|\psi\ra^{ABC})<{E}_{g'',\min}^{(2)}(|\psi\ra^{C|AB})\\
&=&{E}_{g'',\min}^{(2)}(|\psi\ra^{B_2}C)={E}_{g'',\min}^{(2)}(\rho^{BC}),
\eeax
\beax&&{\hat{N}}_{g''}^{(3)}(|\psi\ra^{ABC})<{\hat{N}}_{g''}^{(2)}(|\psi\ra^{A|BC})\\
&=&{\hat{N}}_{g''}^{(2)}(|\psi\ra^{AB_1})={\hat{N}}_{g''}^{(2)}(\rho^{AB}), 
\eeax
\beax&&{\hat{N}}_{g''}^{(3)}(|\psi\ra^{ABC})<{\hat{N}}_{g''}^{(2)}(|\psi\ra^{AB|C})\\
&=&{\hat{N}}_{g''}^{(2)}(|\psi\ra^{B_2C})={\hat{N}}_{g''}^{(2)}(\rho^{BC}).
\eeax
In addition,
\beax
C(\rho^{BD})\approx0.839>C_{gme}(|\xi\ra)
\eeax
for the pure state $|\psi\ra$ in Eq.~\eqref{eg3}. Let
\bea\label{eg4}
|\zeta\rangle^{ABC} =\lambda_0|000\rangle+\lambda_2|101\rangle+\lambda_3| 110\rangle
\eea
with $\lambda_0\geq\lambda_2\geq\lambda_3>0$. If we take $\lambda_0=\frac{\sqrt{5}}{\sqrt{12}}$, $\lambda_2=\frac{1}{\sqrt{3}}$, and $\lambda_3=\frac{1}{2}$ in Eq.~\eqref{eg4}, then $E_{g'',2}^{(3)}(|\zeta\ra^{ABC})=1/4$, but $E_{g'',2}^{(2)}(|\zeta\ra^{A|BC})=5/12$, $E_{g'',2}^{(2)}(|\zeta\ra^{AB|C})=1/3$. In general, for the state
\beax 
|\omega\rangle^{ABC} =\lambda_{0}|000\rangle+\lambda_{2}|101\rangle
+\lambda_{3}| 110\rangle+\lambda_{4}|111\rangle
\eeax
with $\lambda_0\lambda_4>0$, $\max\{\lambda_2, \lambda_3\}>0$ and $\min\{\lambda_2, \lambda_3\}=0$,
then (i) $\rho^{AC}$ and $\rho^{BC}$ are separable while $\rho^{AB}$ is entangled whenever $\lambda_3>0$,
and (ii) $\rho^{AB}$ and $\rho^{BC}$ are separable while $\rho^{AC}$ is entangled whenever $\lambda_2>0$.
From this we can arrive at (i) if $\lambda_{4}$ is small enough, then
\beax
C_{gme}(|\omega\rangle^{ABC})=C(|\omega\rangle^{AB|C})<C(|\omega\rangle^{A|BC}),\quad\\ C(|\omega\rangle^{AB|C})<C(|\omega\rangle^{B|AC}),\quad\quad\quad\quad\quad\\ C_{gme}(|\omega\rangle^{ABC})<C(\rho^{AB}),~\quad\quad\quad\quad\quad
\eeax
and (ii) if $\lambda_{4}$ is small enough, then \beax
C_{gme}(|\omega\rangle^{ABC})=C(|\omega\rangle^{B|AC})<C(|\omega\rangle^{C|AB}),\quad\\ C(|\omega\rangle^{B|AC})<C(|\omega\rangle^{A|BC}),\quad\quad\quad\quad\quad\\ C_{gme}(|\omega\rangle^{ABC})<C(\rho^{AC}).~\quad\quad\quad\quad\quad
\eeax
For example, when taking $\lambda_0^2=7/9$, $\lambda_3=\lambda_4=1/3$,
we get 
\beax C_{gme}(|\omega\rangle^{ABC})\approx0.5879,\quad\quad~~\\
 C(|\omega\rangle^{A|BC})=C(|\omega\ra^{B|AC})\approx0.8315, \\
 C(\rho^{AB})\approx0.8090;\quad\quad\quad\quad
 \eeax 
when taking $\lambda_0^2=7/9$, $\lambda_2=\lambda_4=1/3$,
we get 
\beax C_{gme}(|\omega\rangle^{ABC})\approx0.5879,\quad\quad~~\\
 C(|\omega\rangle^{A|BC})=C(|\omega\ra^{C|AB})\approx0.8315,\\
  C(\rho^{AC})\approx0.8090.\quad\quad\quad\quad
\eeax

\begin{figure}[htbp]	
	\centering
	\subfigure[$C_g$]{
		\hspace{-28mm}\begin{minipage}[c]{0.3\textwidth}
			\centering
			\includegraphics[width=78mm]{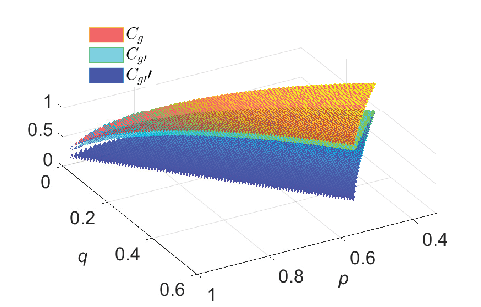}
		\end{minipage}
	}	\\
	\subfigure[$E_{g,\mathcal{F}'}$]{
		\hspace{-28mm}\begin{minipage}[c]{0.3\textwidth}
			\centering
			\includegraphics[width=78mm]{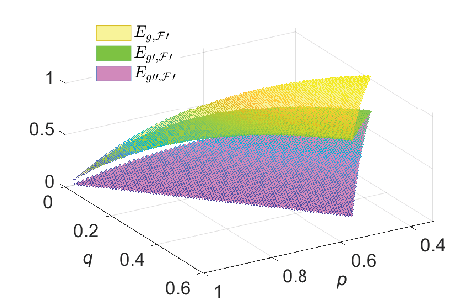}
		\end{minipage}
	}	 \\
	
	\subfigure[$E_{g,2}$]{
		\hspace{-28mm}\begin{minipage}[c]{0.3\textwidth}
			\centering
			\includegraphics[width=78mm]{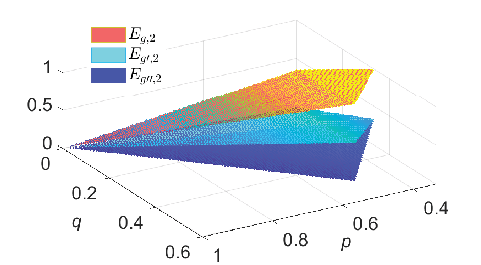}
		\end{minipage}
	}	
	\caption{\label{fig2}(color online). Comparing (a) $C_g^{(3)}$, $C_{g'}^{(3)}$ and $C_{g''}^{(3)}$, (b) $E_{g,\mathcal{F}'}^{(3)}$, $E_{{g',\mathcal{F}'}}^{(3)}$ and $E_{g'',\mathcal{F}'}^{(3)}$, (c) $E_{g,2}^{(3)}$, $E_{g',2}^{(3)}$ and $E_{g'',2}^{(3)}$ for $|\Phi\ra$ with $p\geq q\geq 1-p-q>0$, respectively.}
\end{figure}

For the generalized GHZ state
\bea
|GHZ\ra=\lambda_0|0\ra^{\ot n}+\lambda_1|1\ra^{\ot n}+\cdots\lambda_{d-1}|d-1\ra^{\ot n},~~~
\eea
$E_{g''}^{(n)}$ and $\mE_{g''}^{(n)}$ are complete monogamous and tightly complete monogamous. For this state, $E_{g''}^{(n)}=E_{g'}^{(n)}=\mE_{g''}^{(n)}=\mE_{g'}^{(n)}$, and $nE_{g''}^{(n)}=nE_{g'}^{(n)}=2E_{g}^{(n)}$. Moreover, for such a state, all the entanglement are shared between all of the particles. We thus regard this state as the maximal genuinely entangled state, and it reaches the maximal value whenever $\lambda_0=\lambda_1=\cdots=\lambda_{d-1}={1}/{\sqrt{d}}$ for the multi-qudit case.

Comparing $E_{g''}^{(3)}$ with $E_{g'}^{(3)}$ and $E_{g}^{(3)}$, $E_{g'}^{(3)}$ seems the best one since (i) it is complete and completely monogamous whenever the reduced function is strictly concave, (ii) it can be easily calculated, and (iii) it is monogamous iff it is completely monogamous. For the case of $n\geq4$, $E^{(n)}$, $\mE^{(n)}$, $E_g^{(n)}$, and $\mE_g^{(n)}$ seems better than the other cases as a MEM/GMEM as these measures admit the postulates of a complete MEM/GMEM.

At last, we calculate these GMEMs for the following examples,
\beax
|\Psi\ra&=&\sqrt{t}|000\ra+\sqrt{1-t}|111\ra,\\
|\Phi\ra&=&\sqrt{p}|100\ra+\sqrt{q}|010\ra+\sqrt{1-p-q}|001\ra.
\eeax
For the GHZ class state $|\Psi\ra$, $E_{g'}$ coincides with $E_{g''}$ and $E_{g'}$ is equivalent to $E_g$ (see Fig.~\ref{fig1} for detail). For $|\Phi\ra$, $E_g$, $E_{g'}$ and $E_{g''}$ reflect roughly the same tendency (see Fig.~\ref{fig2} for detail).

\section{Conclusion}

We developed a grained scenario of investigating the MEM and GMEM based on its reduced functions and then explored these
measures in light of the framework of the complete MEM and complete monogamy relation respectively. We provided criteria that can verify whether a MEM/GMEM is good or not. By comparison, with the assumption that the maximal reduced function does not increase on average under LOCC, for tripartite case, the MEM and GMEM via the maximal reduced function seems finer than that of the minimal reduced function as it not only can be easily calculated but also is complete and completely monogamous. And for the $n$-partite case with $n\geq4$, the MEM and GMEM via the sum of the reduced function sound better than the other one in the framework of complete MEM and complete monogamy relation.

In addition, our findings show that, whether the reduced function is strictly concave and whether it is subadditive
is of crucial important. We can also conclude that the monogamy is stronger than the complete monogamy in general, they are equivalent to each other for some case such as the MEM and GMEM via the maximal reduced function for the tripartite case, and the tightly complete monogamy is stronger than the complete monogamy in general. We also find that, in the framework of complete MEM, the hierarchy condition is stronger than the unification condition in general but it is not true for some case such as the MEM and GMEM via the maximal bipartite entanglement.

\begin{acknowledgements}
This work is supported by the National Natural Science Foundation of China under Grant No.~11971277 and the High-Level Talent Research Start-up Fund of Inner Mongolia University under Grant No. 10000-2311210/049.
\end{acknowledgements}




\end{document}